\documentclass[]{article}
\pdfoutput=1

\usepackage[utf8]{inputenc}
\usepackage{fullpage} 
\usepackage{breakcites}
\usepackage{graphicx}
\usepackage{amssymb}
\usepackage{epstopdf}
\usepackage{xspace} 
\usepackage[usenames,dvipsnames]{color}
\usepackage{hyperref}
\usepackage{wrapfig}
\usepackage{ctable}
\usepackage{xcolor}
\usepackage{colortbl}
\usepackage{booktabs}
\usepackage{todonotes}
\usepackage{array}
\usepackage{subcaption}

\newcommand{\ie}{i.\,e.\xspace}
\newcommand{\eg}{e.\,g.\xspace}
\newcommand{\etal}{et al.\xspace}

\newcommand{\nwk}{\textsf{NetworKit}\xspace}

\definecolor{bg}{rgb}{0.985,0.985,0.985}

\hypersetup{breaklinks=true,
            pdfauthor={Christian L. Staudt, Aleksejs Sazonovs, Henning Meyerhenke},
            pdftitle={NetworKit: A Tool Suite for Large-scale Network Analysis},
            colorlinks=true,
            citecolor=PineGreen,
            urlcolor=MidnightBlue,
            linkcolor=PineGreen,
            pdfborder={0 0 0}}
\definecolor{bg}{rgb}{0.985,0.985,0.985}

\title{NetworKit: \\ A Tool Suite for Large-scale Complex Network Analysis}
\author{Christian L. Staudt, Aleksejs Sazonovs, Henning Meyerhenke }
\date{}                                           

\begin{document}
\maketitle

\begin{abstract} \small\baselineskip=9pt 

We introduce NetworKit, an open-source software package for analyzing the structure of large complex networks. Appropriate
algorithmic solutions are required to handle increasingly common large graph
data sets containing up to billions of connections. 
We describe the methodology applied to develop scalable solutions to network analysis problems, including techniques like parallelization, heuristics for computationally expensive problems, efficient data structures, and modular software architecture.
Our goal for the software is
to package results of our algorithm engineering efforts and put them
into the hands of domain experts. 
NetworKit is implemented as a hybrid combining the
kernels written in C++ with a Python
front end, 
enabling integration into the Python ecosystem of tested tools for data analysis and
scientific computing. 
The package provides a wide range of functionality (including common and novel analytics algorithms and graph generators) and does so via a convenient interface.
In an experimental comparison with related software, NetworKit shows the best performance on a range of typical analysis tasks.

~\\[0.5ex]
\end{abstract}


\section{Motivation}\label{motivation}

A great variety of phenomena and systems have been successfully modeled as complex networks \cite{costa2011analyzing,boccaletti2006complex}.
Accordingly, network analysis methods are quickly becoming pervasive in science, technology and society. 
On a closer look, the rallying cry of the emerging field of network science ("networks are everywhere") is hardly surprising: What is being developed is a set of general methods for the statistics of relational data.
Since promising large network data sets are increasingly common in the age of big data, it is an active current research project to develop scalable methods for the analysis of large networks.
In order to process massive graphs, we need algorithms whose running time is essentially linear in the number of edges.
Many analysis methods have been pioneered on small networks (\eg for the study of social networks prior to the arrival of massive online social networking services), so that underlying algorithms with higher complexity were viable. 
As we shall see in the following, developing a scalable analysis tool suite often entails replacing them with suitable linear- or nearly-linear-time variants.
Furthermore, solutions should employ parallel processing:
While sequential performance is stalling, multicore machines become pervasive, and algorithms and software need to follow this development. 
Within the \nwk project, scalable network analysis methods are developed, tested and packaged as ready-to-use software. 
In this process we frequently apply the following algorithm and software engineering patterns: \textit{parallelization}; \textit{heuristics} or \textit{approximation algorithms} for computationally intensive problems; efficient \textit{data structures}; and \textit{modular} software architecture.
With \nwk, we intend to push the boundaries of what can be done interactively 
on a shared-memory parallel computer, also by users without in-depth programming skills.
The tools we provide make it easy to characterize large networks and are geared towards network science research.

In this work we give an introduction to the tool suite 
and describe the methodology applied during development in terms of algorithm and software engineering aspects.
We discuss methods to arrive at highly scalable solutions to common network analysis problems (Sections~\ref{sec:methodology} and~\ref{sec:patterns}), describe the set of functionality (Sections~\ref{sec:analytics} and~\ref{sec:generators}), present example use cases (Section~\ref{sec:usecases}), compare with related software (Section~\ref{sec:comparison}), and evaluate the performance of analysis kernels experimentally (Section~\ref{sec:performance}).
Our experiments show that \nwk is capable of quickly processing large-scale 
networks for a variety of analytics kernels, and does so faster and with a lower memory footprint than closely related software. 
 We recommend \nwk for the comprehensive structural analysis of massive complex networks (their size is
primarily limited by the available memory).
To this end, a new frontend supports exploratory data analysis with fast graphical reports on structural features of the network (Section~\ref{sec:exploratory}).

\section{Methodology}
\label{sec:methodology}
\subsection{Design Goals.}
There is a variety of software packages which provide graph algorithms in general and network analysis capabilities in particular (see Section~\ref{sec:comparison} for a comparison to related packages). However, \nwk aims to balance a specific combination of strengths. Our software is designed to 
stand out with respect to the following areas:

\emph{Performance.}
Algorithms and data structures are selected and implemented with high performance and parallelism in mind. 
Some implementations are among the fastest in published research. For example, community detection in a $3.3$ billion edge web graph can be 
performed on a 16-core server with hyperthreading in less than three minutes~\cite{staudt2015engineering}.

\emph{Usability and Itegration.}
 Networks are as diverse as the series of questions we might ask of them -- \eg, what is the largest connected component, what are 
the most central nodes in it and how do they connect to each other? A practical tool for network analysis should therefore provide modular functions which do not restrict the user to predefined workflows.
An interactive shell, which the Python language provides, is one prerequisite for that. While \nwk works with the standard Python 3 interpreter, 
calling the module from the \textsf{IPython} shell and \textsf{Jupyter Notebook} HTML interface~\cite{perez2013open} allows us to integrate it into a fully fledged 
 computing environment for scientific workflows, from data preparation to creating figures. It is also easy to set up and control a remote compute server.
As a Python module, \nwk enables seamless integration with Python libraries for scientific computing and data analysis, \eg \textsf{pandas} for 
data frame processing and analytics, \textsf{matplotlib} for plotting or \textsf{numpy} 
and \textsf{scipy} for numerical and scientific computing.
For certain tasks, we provide interfaces to specialized external tools, \eg \textsf{Gephi}~\cite{bastian2009gephi} for graph visualization.

\subsection{Architecture.}\label{architecture}

\begin{figure}[!h]
  \begin{center}
\includegraphics[width=.8\textwidth]{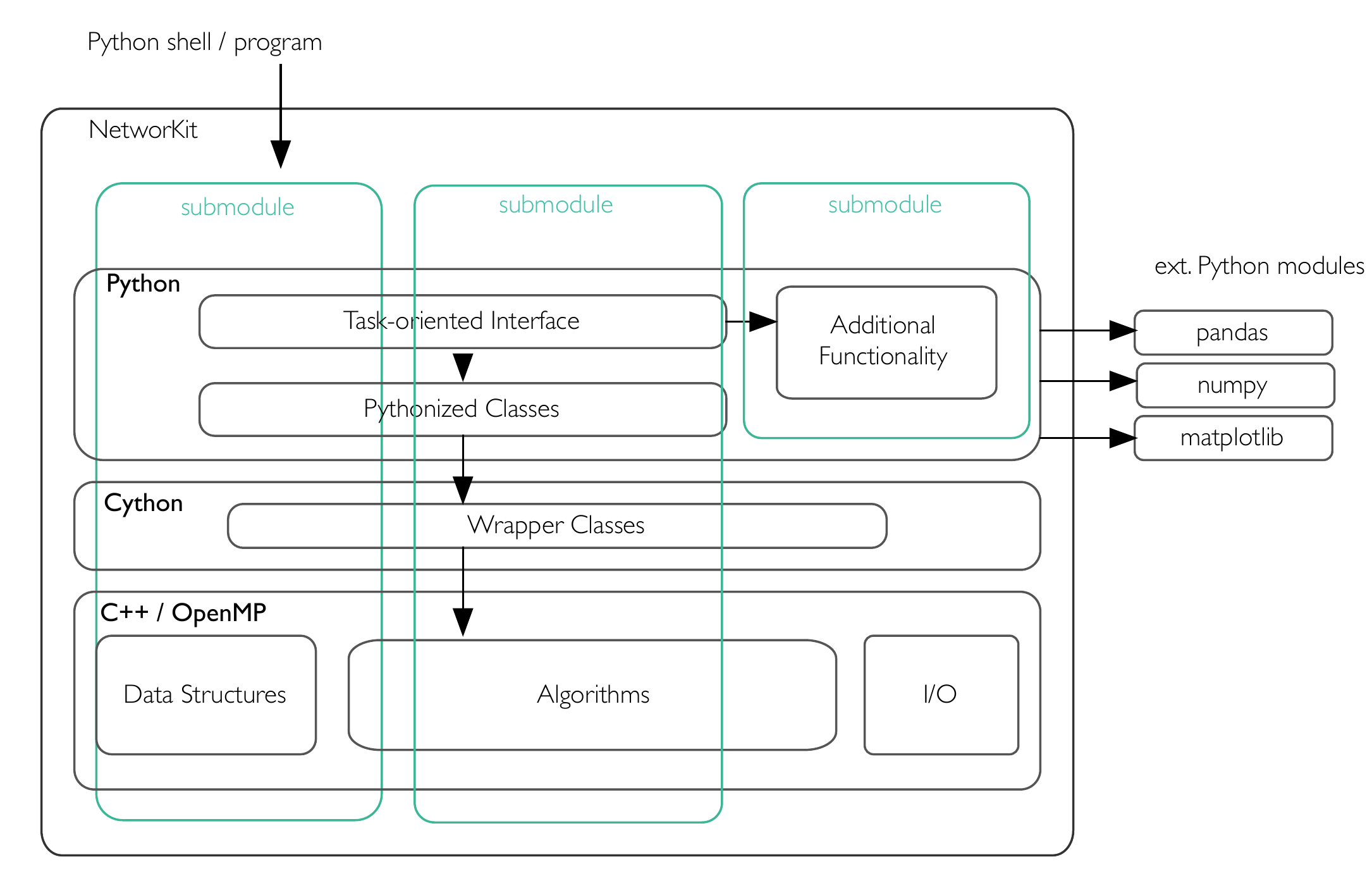}
\caption{\nwk architecture overview ($\rightarrow$ represents call from/to)}
\label{fig:architecture}
  \end{center}
\end{figure}

In order to achieve the design goals described above, we implement \nwk as a two-layer hybrid of performance-aware code written in C++ with an 
interface and additional functionality written in Python. \nwk is distributed as a Python package, ready to be used interactively from a Python shell, which is the main usage scenario we envision for domain scientists. 
The code can be used as a library for application programming as well, either at the 
Python or C++ level.
Throughout the project we use object-oriented and functional concepts. Shared-memory parallelism is realized with \href{http://openmp.org}{\textsf{OpenMP}}, 
providing loop parallelization and synchronization constructs while abstracting away the details of thread creation and handling. The roughly 45\,000 lines of C++ code include core implementations and unit tests. 
As illustrated in Figure~\ref{fig:architecture},
connecting these native implementations to the Python world is enabled by the \href{http://cython.org/}{Cython} toolchain~
\cite{behnel2011cython}. 
Currently we use Cython to integrate native 
code by compiling it into a Python extension module. 
The Python layer comprises about 4\,000 lines of code.
The resulting Python module \texttt{networkit} is organized into several submodules for different areas of functionality, such as community detection or node centrality. A submodule may bundle and expose C++ classes or exist entirely on the Python layer.

\subsection{Framework Foundations.}\label{features}

As the central data structure, the \texttt{Graph} class implements a directed or undirected, optionally weighted graph using an adjacency array data 
structure with $O(n +m)$ memory requirement for a graph with $n$ nodes and $m$ edges.
Nodes are represented by 64 bit integer indices from a consecutive range, and an 
edge is identified by a pair of nodes. Optionally, edges can be indexed as well.
This approach enables a lean graph data structure, while also allowing arbitrary node and edge attributes to be stored in any container addressable by indices. While some algorithms may benefit from different data layouts,
this lean, general-purpose representation has proven suitable for writing performant implementations.
In particular, it supports dynamic modifications to the graph in a flexible manner, unlike the \textit{compressed sparse row} format common in high-performance scientific computing.
Our graph API aims for an intuitive and concise formulation of graph algorithms on both the C++ and Python layer (see Fig.~\ref{code:betweenness} for an example).
In general, a combinatorial view on graphs -- representing edges as tuples of nodes -- is used.
However, part of \nwk is an algebraic interface that enables the implementation of graph algorithms in terms of various matrices describing the graph, while transparently using the same graph data structure.

\section{Algorithm and Implementation Patterns}
\label{sec:patterns}
As explained in Section~\ref{motivation}, our main focus are scalable algorithms in order
to support network analysis on massive networks. We identify several algorithm
and implementation patterns that help to achieve this goal and present them below by means of case
studies. For experimental results we express processing speed in "edges per second", an intuitive way to aggregate real running time over a set of graphs and normalize by graph size.

\subsection{Parallelism}
Our first case study concerns the \textit{core decomposition} of a graph, which allows a fine-grained subdivision of the node set according to connectedness.  
More formally, the $k$-core is the maximal induced subgraph whose nodes have at least degree $k$.
The decomposition also categorizes nodes according to the highest-order 
core in which they are contained, assigning a \textit{core number} to each node (the largest $k$ for which
the node belongs to the $k$-core).
The sequential kernel implemented in \nwk runs in $O(m)$ time, matching other 
implementations~\cite{Batagelj:2011fk}.
The main algorithmic idea we reuse for computing the core numbers is to start with $k=0$ and increase $k$
iteratively. Within each iteration phase, all nodes with degree $k$ are successively removed (thus,
also nodes whose degree was larger at the beginning of the phase can become affected by a removal
of a neighbor). Our implementation uses a bucket priority queue. From this data structure we can extract
the nodes with a certain minimum residual degree in amortized constant time. The same time holds for updates of the
neighbor degrees, resulting in $O(m)$ in total.

While the above implementation already scales to large inputs, it can still make a significant difference if a user needs to
wait minutes or seconds for an answer. Thus, we also provide a parallel implementation. The
sequential algorithm cannot be made parallel easily due to its sequential access to the bucket priority queue.
For achieving a higher degree of parallelism, we follow~\cite{DBLP:conf/bigdataconf/DasariRZ14}.
Their \textsf{ParK} algorithm replaces the extract-min operation in the above algorithm by identifying the 
node set $V'$ with nodes of minimum residual degree while iterating in parallel over all (active) nodes. $V'$ is then further processed
similarly to the node retrieved by extract-min in the above algorithm, only in parallel again.
\textsf{ParK} thus performs more sequential work, but with thread-local buffers
it relies on a minimal amount of synchronization. Moreover, its data access pattern is more cache-friendly,
which additionally contributes to better performance.

\begin{figure}[!h]
\begin{center}
\includegraphics[width=.75\columnwidth]{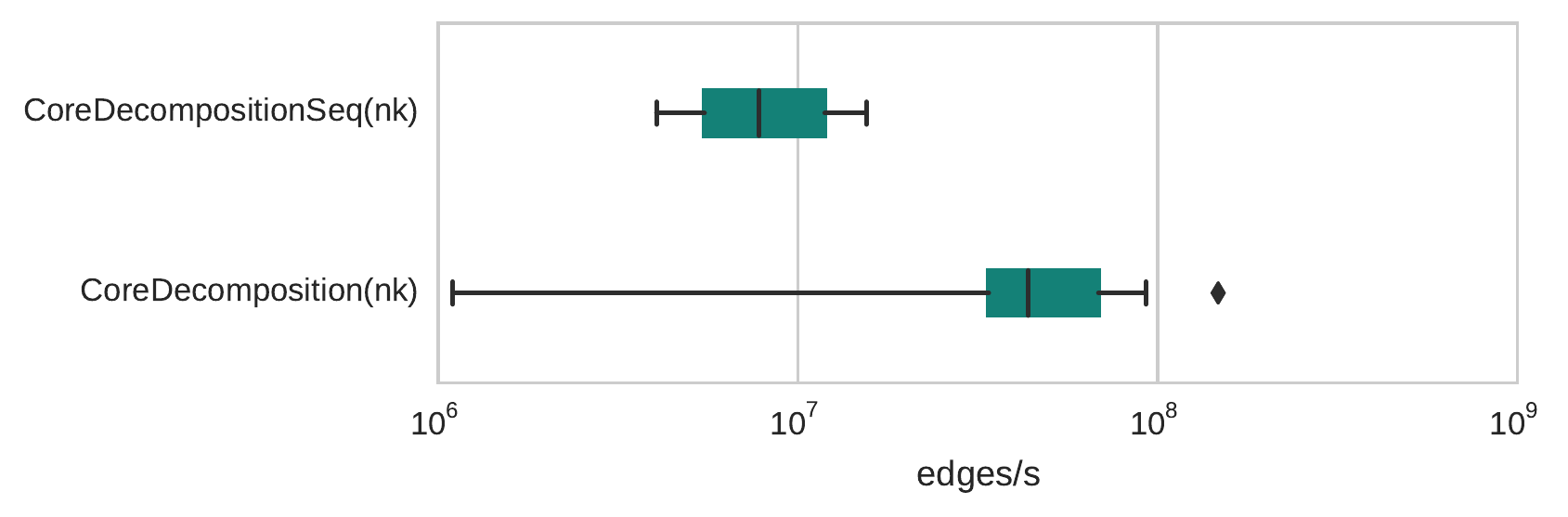}
\end{center}
\caption{Core decomposition: sequential versus parallel performance} 
\label{fig:coreDec}
\end{figure}

Fig.~\ref{fig:coreDec} is the result of running time measurements on a test set of networks (see Sec.~\ref{sec:performance} for the setup). 
We see that on average, processing speed is increased by almost an order of magnitude through parallelization. Some overhead of the parallel algorithm implies that speedup is only noticeable on large graphs, hence the large variance. For example, processing time for the 260 million edge \textsf{uk-2002} web graph is reduced from 22 to 2 seconds.

\subsection{Heuristics and Approximation Algorithms}
\label{sec:heuristics}

In this example we illustrate how inexact methods deliver appropriate solutions for an otherwise computationally impractical problem. 
\textit{Betweenness centrality} is a well-known node centrality measure that has an intuitive interpretation in transport networks:
Assuming that the transport processes taking place in the network are efficient, they follow shortest paths through the network, and therefore preferably pass through nodes with high betweenness. For instance, their removal would interfere strongly with the function of the network. It is clear that network analysts would like to be able to identify such nodes in networks of any size.
\nwk comes with an implementation of the currently fastest known algorithm for betweenness~\cite{Bra01}, which has $O(nm)$ running time in unweighted graphs. 

\begin{figure}[!h]
\begin{center}
\includegraphics[]{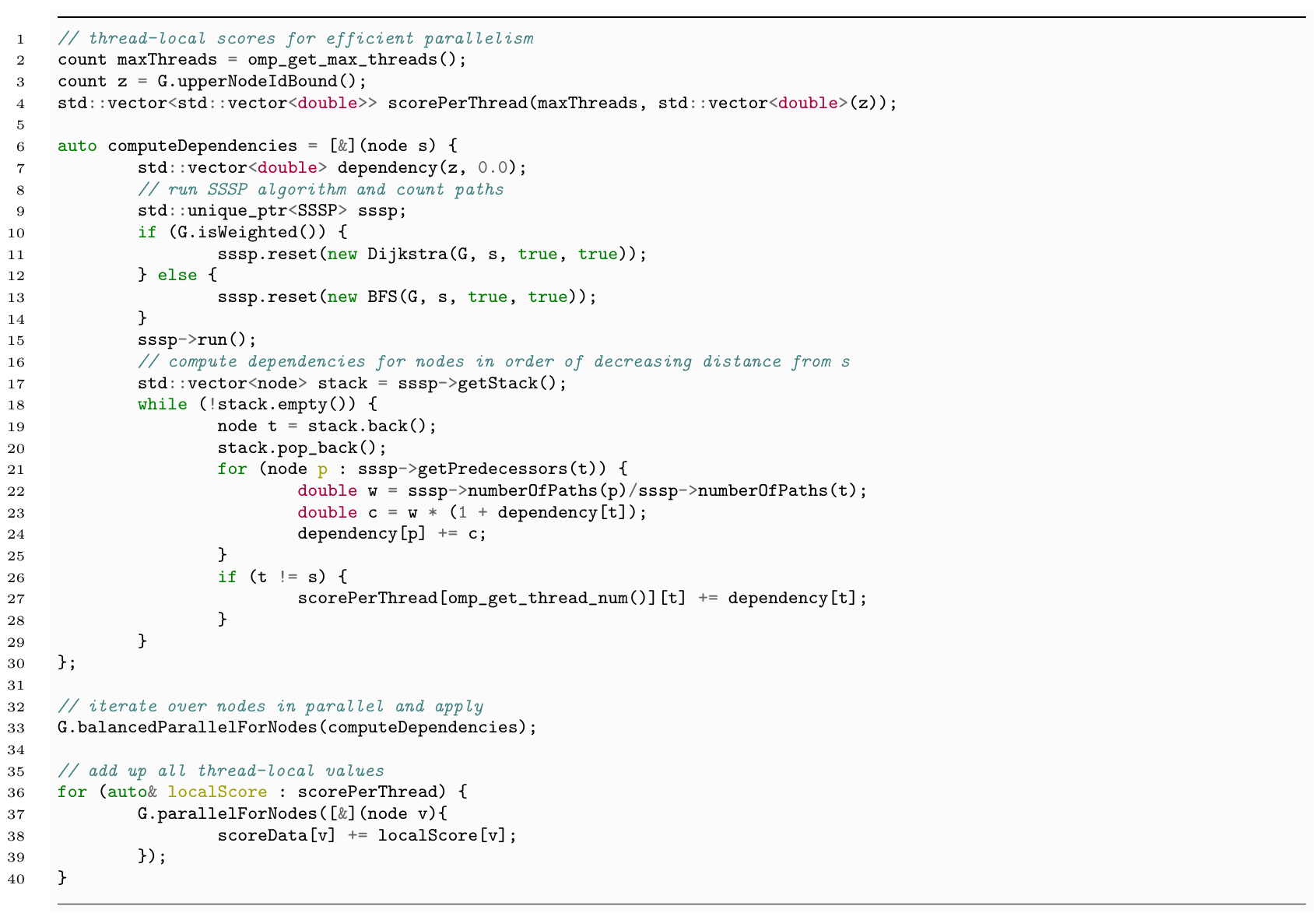}
\end{center}
\caption{Code example: Parallel calculation of betweenness centrality}
\label{code:betweenness}
\end{figure}

With a closer look at the algorithm, opportunities for parallelization are apparent:
Several single-source shortest path searches can be run in parallel to compute the intermediate \textit{dependency} values whose sum yields a node's betweenness.
Figure~\ref{code:betweenness} shows C++ code for the parallel version, which is simplified to focus on the core algorithm, but the actual implementation is similarly concise.
To avoid race conditions, each thread works on its own dependency array, which need to be aggregated into one betweenness array in the end (lines 35-39).

\begin{figure}[!h]
\begin{center}
\includegraphics[width=.75\columnwidth]{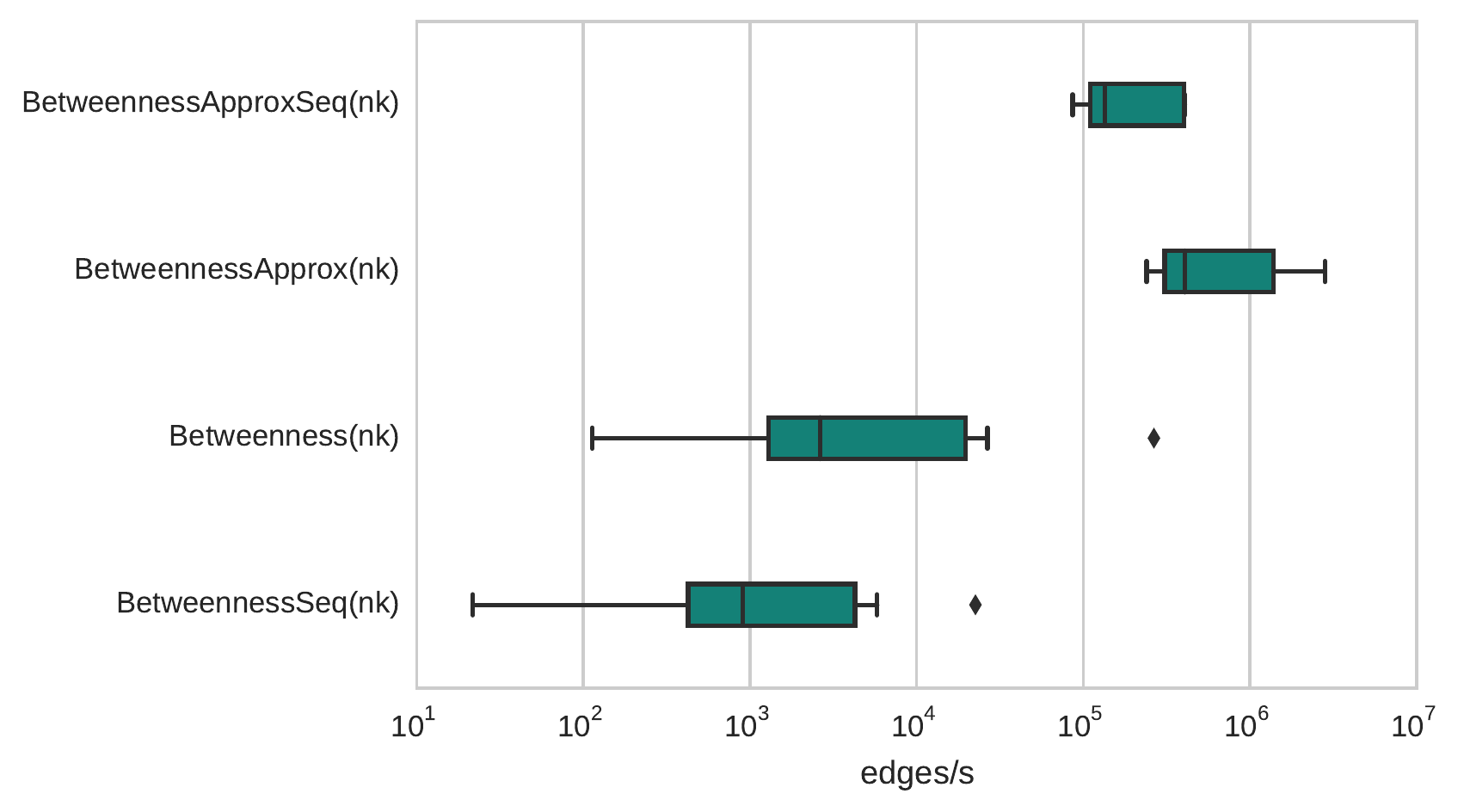}
\end{center}
\caption{Processing speed of exact and inexact algorithms for betweenness centrality} 
\label{fig:bench-betweenness}
\end{figure}

We now evaluate the performance of the implementations experimentally (see Section~\ref{sec:performance} for settings).
Figure~\ref{fig:bench-betweenness} shows aggregated running speed over a set of smaller networks (from Table~\ref{tab:networks}).
In practice, this means that the sequential version of Brandes' algorithm (\textsf{BetweennessSeq}) takes almost 8 hours to process the 
600k edge graph \textsf{caidaRouterLevel} (representing internet router-level topology~\cite{caida}).
Parallelism with 32 (hyper)threads (\textsf{Betweenness}) reduces the running time to ca. 90 minutes. 
Still, parallelization does not change the algorithm's inherent complexity. This means that running times rise so steeply with the size of the input graph that computing an exact solution to betweenness is not viable on the large networks we want to target.
In typical use cases, obtaining exact values for betweenness is not necessary, though.
An approximate result is likely good enough to appreciate the structure of the network for exploratory analysis, and to identify a set of top betweenness nodes.
Therefore, we use a heuristic approach based on computing a relatively small number of randomly chosen shortest-path trees~\cite{geisberger2008better}.
In contrast to the exact algorithm, running the approximative algorithm with 42 samples takes 6 seconds sequentially. 
Applying this algorithm cuts running time by orders of magnitude, but still yields a ranking of nodes that is highly similar to a ranking by exact betweenness values.
We observe that the distribution of relative rank errors (exact rank divided by approximated rank) has little variance around 1.0. Nodes on average maintain the rank they would have according to exact betweenness even with such a small number of samples. Concretely we see, for instance, that the top ten nodes in the exact betweenness ranking are $(3,14,22,2,58,10,54,39,127,55)$ and $(14,3,22,58,2,55,10,127,26,6)$ in the approximate ranking. 
Experiments of this type (see~\cite{geisberger2008better}) confirm that in typical cases betweenness can be closely approximated with a relatively small number $s$ of shortest-path searches.
Therefore we can replace an $O(n m)$ algorithm with one of time complexity $O(s m)$ in many use cases. 
The inexact algorithm offers the same opportunities for parallelization, yielding additional speedups: In the example above, parallel running time is down to 1.5 seconds on 32 (hyper)threads.

If a true approximation with a guaranteed error bound is desired, \nwk users can apply another inexact algorithm~\cite{Riondato:2015} which accepts an error bound parameter $\epsilon$. It sacrifices some computational efficiency but allows a proof that the resulting betweenness scores have at most $\pm\epsilon$ difference from the exact scores (with a user-supplied probability).

\subsection{Efficient Data Structures}
\label{sub:data-structures}
\begin{wrapfigure}[19]{R}{0.35\linewidth}
\vspace*{-1\baselineskip}
\includegraphics[scale=0.8]{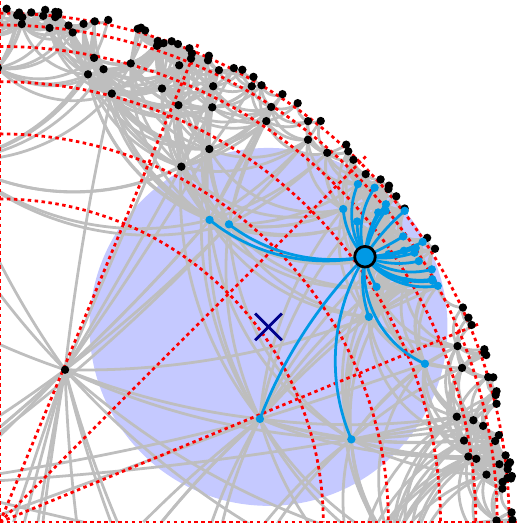}
\caption{Black: Points corresponding to network nodes. Grey: Edges between nearby points.
Red: Boundaries of polar quadtree cells. Blue: Sample node whose neighborhood is visualized
by the purple circle and blue edges.}
\label{fig:graph-polar-quadtree}
\end{wrapfigure}
The case study on data structures deals with 
a \textit{generative network model}. Such models are important as they simplify complex 
network research in several respects (see Section~\ref{sec:generators}).
\emph{Random hyperbolic graphs} (RHGs)~\cite{Krioukov2010} are 
very promising in this context, since theoretical analyses have shown that RHGs have many features also found in real 
complex networks~\cite{bode2014probability,DBLP:conf/icalp/GugelmannPP12,kiwi2015bound}. The model is based on hyperbolic geometry, into 
which complex networks can be embedded naturally.
During the generation process vertices are distributed randomly on a hyperbolic disk of radius $R$ and edges are inserted for every vertex pair whose distance is below $R$.
The straightforward RHG generation process would probe the distance of all pairs, yielding a quadratic time complexity.
This impedes the creation of massive networks. \nwk provides the first generation algorithm 
for RHGs with subquadratic running time ($O((n^{3/2}+m) \log n)$ with high 
probability)~\cite{LoozMP15generating}. 
The acceleration stems primarily from the reduction of distance computations through
a polar quadtree adapted to hyperbolic space. Instead of probing each pair of nodes,
the generator performs for each node one efficient range query supported by the quadtree.
In practice this leads to an acceleration of at least two orders of magnitude. With the quadtree-based
approach networks with billions of edges can be generated in parallel in a few 
minutes~\cite{LoozMP15generating}. By exploiting previous results on efficient Erd\H{o}s-R\'{e}nyi graph 
generation~\cite{batagelj2005efficient},
the quadtree can be extended to use more general neighborhoods~\cite{von2015querying}.

\subsection{Modular Design}

In terms of software design, we aim at a modular architecture with proper encapsulation of algorithms into software components (classes and modules). This requires extensive software engineering work but has clear benefits.
Among them are extensibility and code reuse:  For example, new centrality measures can be easily added by implementing a subclass with the code specific to the centrality computation, while code applicable to all centrality measures and a common interface remains in the base class. Through these and other modularizations, developers can add a new centrality measure and get derived measures almost "for free". 
These include for instance the \textit{centralization index}~\cite{freeman1979centrality} and the \textit{assortativity coefficient}~\cite{freeman1979centrality}, which can be defined with respect to any node centrality measure and may in each case be a key feature of the network.

Modular design also allows for optimizations on one algorithm to benefit other client algorithms.
For instance, betweenness and other centrality measures (such as closeness) require the computation of shortest paths, which is done via breadth-first search in unweighted graphs and Dijkstra's algorithm in weighted graphs,
decoupled to avoid code redundancy (see lines 10-14 in Fig.~\ref{code:betweenness}).

\section{Analytics}
\label{sec:analytics}

%
The following describes the core set of network analysis algorithms implemented in \nwk.
 In addition, \nwk also includes a collection of basic graph algorithms, such as breadth-first 
and depth-first search or Dijkstra's algorithm for shortest paths. Table~\ref{tab:algorithms} summarizes the core set of algorithms for typical problems.

\begin{table}[!h]
\begin{center}
\begin{tabular}{|p{1.5cm}|p{4cm}|p{4cm}|p{2cm}|p{2cm}|}
\hline
\rowcolor{gray!80}
\bf category               & \bf task                         & \bf algorithm  & \bf time & \bf space  \\ \hline
centrality        & degree                       &      --                 & $O(n)$            & $O(n)$            \\ \rowcolor{bg}
                       & betweenness                  &              \cite{Bra01}         &       $O(nm)$          &    $O(n+m)$              \\
                       & ap. betweeenness             &      \cite{geisberger2008better},\cite{Riondato:2015}                 & $O(sm)$           &      $O(n+m)$            \\ \rowcolor{bg}
                       & closeness                    &      shortest-path search from each node               &      $O(mn)$           &           $O(n)$         \\
                       & ap. closeness                &            \cite{eppstein2004fast}           & $O(sm)$           &        $O(n)$          \\ \rowcolor{bg}
                       & PageRank                     & power iteration       & $O(m)$   typical   (Sec.~\ref{sec:centrality})     & $O(n)$             \\
                       & eigenvector centrality       &         power iteration              & $O(m)$   typical         & $O(n)$             \\ \rowcolor{bg}
                       & Katz centrality              &     \cite{katz1953new}                 &       $O(m)$   typical          &       $O(n)$            \\
                       & $k$-path centrality            &    \cite{alahakoon2011k}                    &     \multicolumn{2}{p{2.75cm}|}{see     \cite{alahakoon2011k}}                         \\ \rowcolor{bg}
                       & local clustering coefficient & parallel iterator     &  $O(n d^2)$               &    $O(n)$              \\
                       & $k$-core decomposition         & \cite{DBLP:conf/bigdataconf/DasariRZ14}               &    $O(m)$             &                  \\ \rowcolor{bg}
partitions     & connected components         & BFS                   &    $O(m)$             &        $O(n)$          \\
                       & community detection          & PLM, PLP \cite{staudt2015engineering}            &     $O(m)$            &  $O(m)$,$O(n)$                \\ \rowcolor{bg}
               global        & diameter    & iFub~\cite{crescenzi2013computing}                  &    $O(m)$ typical             &       $O(n)$            \\ \hline

\end{tabular}
\end{center}
\caption{Selection of analysis algorithms contained in \nwk. Complexity expressed in terms of $n$ nodes, $m$ edges, $s$ samples and maximum node degree $d$}
\label{tab:algorithms}
\end{table}

\subsection{Global Network Properties}

Global properties include simple statistics such as the number of nodes and edges and the graph's density, as well as properties related to distances: The \textit{diameter} of a graph is the maximum length of a shortest path between any two nodes.
We use the \textsf{iFUB} algorithm~\cite{crescenzi2013computing} both for the exact computation as well as an estimation of a lower and upper bound on the diameter. \textsf{iFub} has a worst case complexity of $O(nm)$ but has shown excellent typical-case performance on complex networks, where it often converges on the exact value in linear time.

\subsection{Node Centrality}
\label{sec:centrality}

Node centrality measures quantify the structural importance of a node within a network. More precisely, we consider a node centrality measure as any function which assigns to each node an attribute value of (at least) ordinal scale of measurement. The assigned value depends on the position of a node within the network as defined by a set of edges.

The simplest measure that falls under this definition is the \textit{degree}, \ie the number of connections of a node. The distribution of degrees  plays  an important role in characterizing a network.
\textit{Eigenvector centrality} and its variant \textit{PageRank}~\cite{page1999pagerank} assign relative importance to nodes according to their connections, incorporating the idea that edges to high-scoring nodes contribute more.
Both variants are implemented in \nwk based on parallel power iteration, whose convergence time depends on a numerical error tolerance parameter and spectral properties of the network, but is among the fast linear-time algorithms for typical inputs.
For \textit{betweenness centrality} we provide the solutions discussed in Sec.~\ref{sec:heuristics}.
Similar techniques are applied for computing \textit{closeness centrality} exactly and approximately~\cite{eppstein2004fast}.
Our current research extends the former approach to dynamic graph processing~\cite{DBLP:conf/esa/BergaminiM15, DBLP:conf/alenex/BergaminiMS15}.
The \textit{local clustering coefficient} expresses how many of the possible connections between neighbors of a node exist, which can be treated as a node centrality measure according to the definition above~\cite{newman2010networks}.
In addition to a parallel algorithm for custering coefficients, \nwk also implements a sampling approximation algorithm~\cite{schank2005approximating},
whose constant time complexity is independent of graph size.
Given \nwk's modular architecture, further centrality measures can be easily added.

\subsection{Edge Centrality, Sparsification and Link Prediction}
\label{sub:edge-centrality}

The concept of centrality can be extended to edges: Not all edges are equally important for the structure of the network,
and scores can be assigned to edges depending on the graph structure such that they can be ranked  (\eg \textit{edge betweenness}, which depends on the number of shortest paths passing through an edge). 

While such a ranking is illuminating in itself, it can also be used to filter edges and thereby reduce the size of data. 
\nwk includes a wide set of edge ranking methods, with a focus on sparsification techniques meant to preserve certain properties of the network. For instance, we show that a method that ranks edges leading to high-degree nodes (hubs) closely preserves many properties of social networks, including diameter, degree distribution and centrality measures.
Other methods, including a family of \textit{Simmelian backbones}, assign higher importance to edges within dense regions of the graph and hence preserve or emphasize communities.
Details are reported in our recent experimental study~\cite{lindner2015structure}. 
While currently experimental and focused on one application, namely structure-preserving sparsification, the design is extensible so that general edge centrality indices can be easily implemented.

A somewhat related problem, conceptually and through common methods, is the task of \textit{link prediction}. Link prediction algorithms examine the edge structure of a graph to derive similarity scores for unconnected pairs of nodes. Depending on the score, the existence of a future or missing edge is inferred. \nwk includes implementations for a wide variety of methods from the literature~\cite{esders2015linkprediction}.

\subsection{Partitioning the Network}

Another class of analysis methods partitions the set of nodes into subsets depending on the graph structure.
For instance, all nodes in a \textit{connected component} are reachable from each other. A network's connected components can be computed in linear time using breadth-first search.
\textit{Community detection} is the task of identifying groups of nodes in the
network which are significantly more densely connected among each other
than to the rest of nodes. It is a data mining problem where various definitions of the structure to be discovered -- the community -- exist. 
This fuzzy task can be turned into a well-defined though NP-hard optimization problem by using community quality measures, first and foremost 
\textit{modularity}~\cite{girvan2002community}. 
We approach community detection from the perspective of modularity maximization and engineer parallel heuristics which deliver a good tradeoff between solution quality 
and running time~\cite{staudt2015engineering}. 
The \textsf{PLP} algorithm implements community detection by label propagation~\cite{Raghavan:2007fk}, which extracts communities from a labelling of the node set. 
The \emph{Louvain method} for community detection~\cite{Blondel:2008uq} can be classified as
a locally greedy, bottom-up multilevel algorithm.
We recommend the \textsf{PLM} algorithm with optional refinement step as the default choice for modularity-driven community detection in large networks.
For very large networks in 
the range of billions of edges, \textsf{PLP} delivers a better time to solution, albeit with a qualitatively different solution and worse modularity.

\section{Network Generators}
\label{sec:generators}

\begin{table}[!h]
\centering
\begin{tabular}{|p{8cm}|p{7cm}|}
\rowcolor{gray!80}
\hline
\bf model [and algorithm] & \bf description \\
\hline 
\textit{Erd\H{o}s-R\'{e}nyi}~\cite{P.Erdos} [\cite{batagelj2005efficient}] & random edges with uniform probability \\ \rowcolor{bg}
\textit{planted partition / stochastic blockmodel} & dense areas with sparse connections\\ 
\textit{Barabasi-Albert}~\cite{albert2002statistical} & preferential attachment process resulting in power-law 
degree distribution\\ \rowcolor{bg}
\textit{Recursive Matrix (R-MAT)} \cite{chakrabarti2004r} & power-law degree distribution, small-world property, self-similarity \\
\textit{Chung-Lu}~\cite{aiello2000random} &  replicate a given degree distribution\\ \rowcolor{bg}
\textit{Havel-Hakimi}~\cite{hakimi1962realizability} & replicate a given degree distribution \\
\textit{hyperbolic unit-disk model}~\cite{Krioukov2010} [\cite{LoozMP15generating}] & large networks, power-law degree distribution and high clustering \\ \rowcolor{bg}
\textit{LFR} \cite{lancichinetti2009benchmarks} & complex networks containing communities \\
\hline
\end{tabular}
\caption{Overview of network generators}
\label{tab:generators}
\end{table}

\textit{Generative network models} aim to explain how networks form and evolve specific structural features. Such models and their implementations as 
generators have at least two important uses: On the one hand, algorithm or software engineers want generators for synthetic datasets which can be 
arbitrarily scaled and parametrized and produce graphs which resemble the real application data. On the other hand, network scientists employ
models to increase their understanding of network phenomena. \nwk provides a versatile collection of graph generators for this purpose, summarized in Table~\ref{tab:generators}.

\section{Example Use Cases.}
\label{sec:usecases}
%
In the following, we present possible workflows and use cases, highlighting the capabilities of \nwk as a data analysis tool and a library.

\begin{figure}[!h]
\begin{center}
\includegraphics[width=0.5\columnwidth]{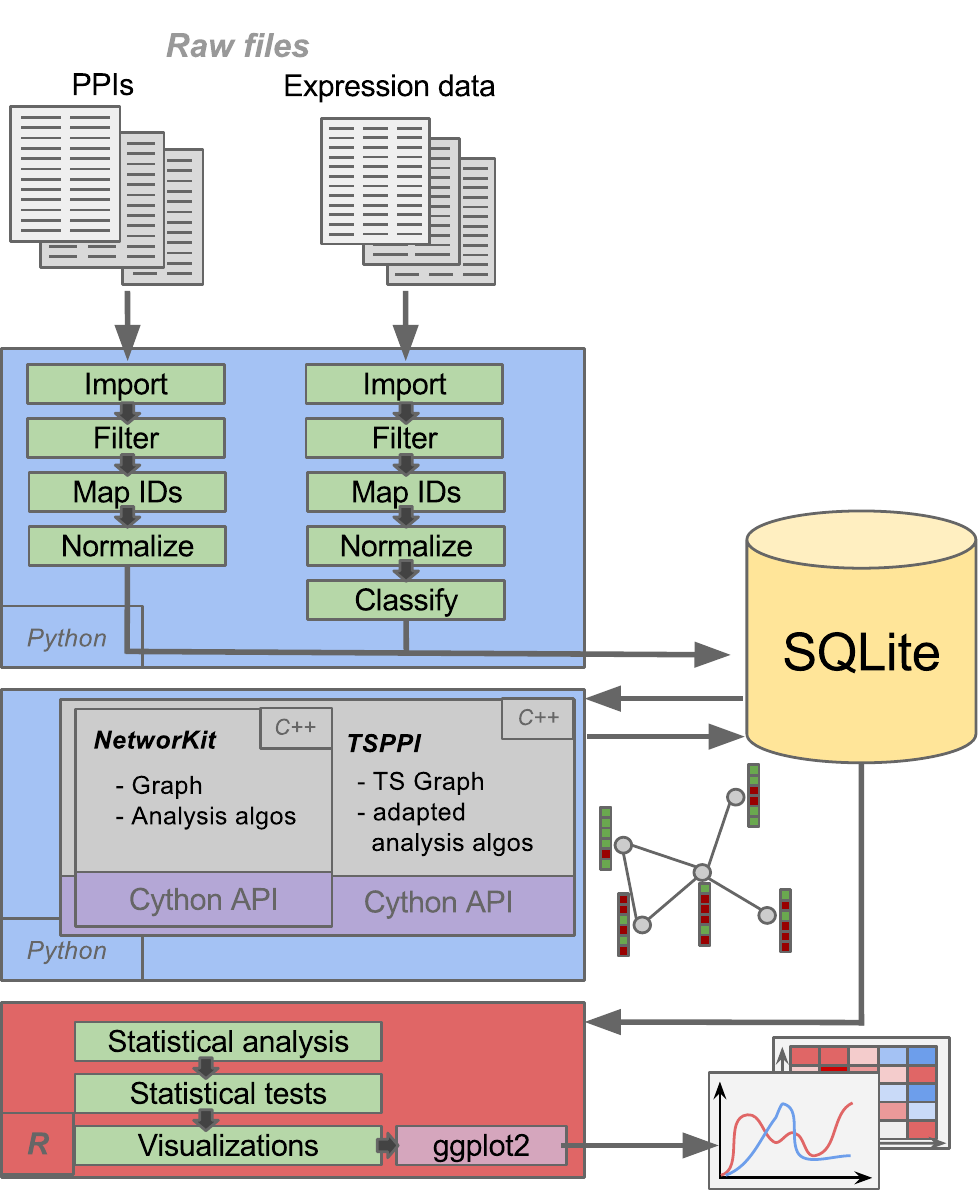}
\caption{PPI network analysis pipeline with \nwk as central component}
\label{fig:pipeline}
\end{center}
\end{figure}

\subsection{As a Library in an Analysis Pipeline}

A master's thesis~\cite{flickanalysis} provides an early example of \nwk as a component in an application-specific data mining pipeline (Fig.~\ref{fig:pipeline}). This pipeline performs analysis of \textit{protein-interaction (PPI) networks}.
and implements a preprocessing stage in Python, in which networks are compiled from heterogeneous data sets containing interaction data as well as \textit{expression data} about the occurrence of proteins in different cell types.
 During the network analysis stage, preprocessed networks are retrieved from a database, and \nwk is called via the Python frontend.
The C++ core has been extended to enable more efficient analysis of
\textit{tissue-specific} PPI networks, by implementing in-place filtering of the network to the subgraphs of proteins that occur in given cell types.
Finally, statistical analysis and visualization is applied to the network analysis data.
The system is close to how we envision \nwk as a high-performance algorithmic component in a real-world data analysis scenario, 
and we therefore place emphasis on the toolkit being easily scriptable and extensible.

\subsection{Exploratory Network Analysis with Network Profiles}
\label{sec:exploratory}

\begin{figure}[!h]
\begin{center}
\includegraphics[width=.8\columnwidth]{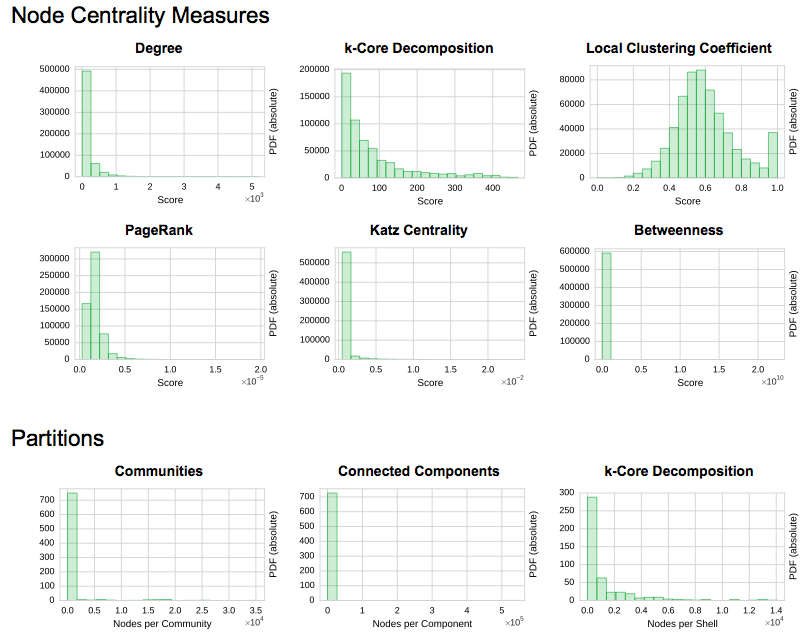}
\caption{Overview on the distributions of node centrality measures and size distributions of different network partitions -- here: a Facebook social network}
\label{fig:measure-overview}
\end{center}
\end{figure}

\begin{figure}[!h]
\begin{center}
\includegraphics[width=\columnwidth]{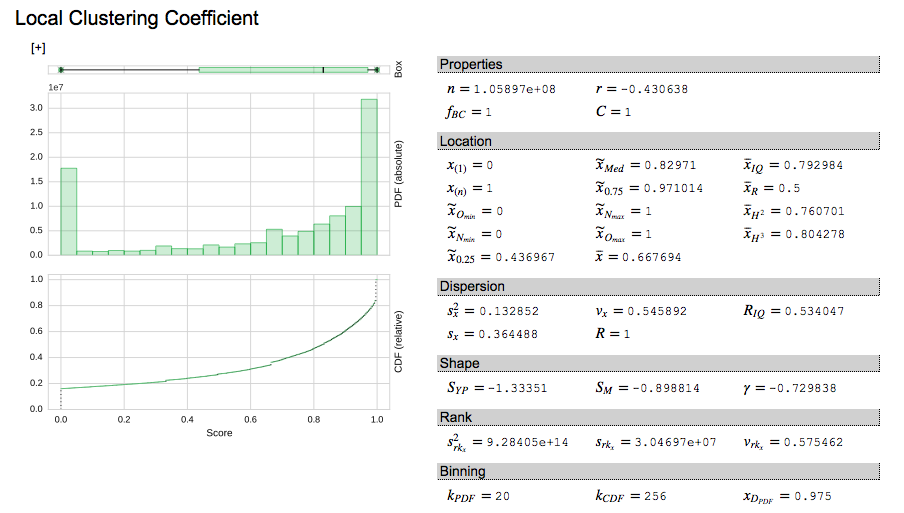}
\caption{Detailed view on the distribution of node centrality scores -- here: local clustering coefficients of the 3 billion edge web graph \textsf{uk-2007}} 
\label{fig:centrality-detail}
\end{center}
\end{figure}

\begin{figure}[!h]
\begin{center}
	\begin{subfigure}[b]{0.45\textwidth}
	\includegraphics[width=\columnwidth]{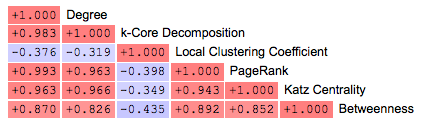}
	\caption{-- in the social network \textsf{fb-Texas84}}
	\label{fig:facebook-correlations}
	\end{subfigure} 
	\quad
	\begin{subfigure}[b]{0.45\textwidth}
	\includegraphics[width=\columnwidth]{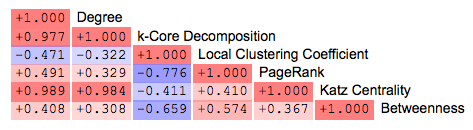}
	\caption{-- in the connectome network \textsf{con-fiber\_big}}
	\label{fig:connectome-correlations}
	\end{subfigure}
\end{center}
\caption{Correlation between node centrality measures --}
\end{figure}

\begin{figure}[!h]
\begin{center}
	\begin{subfigure}[b]{0.4\textwidth}
	\includegraphics[width=\columnwidth]{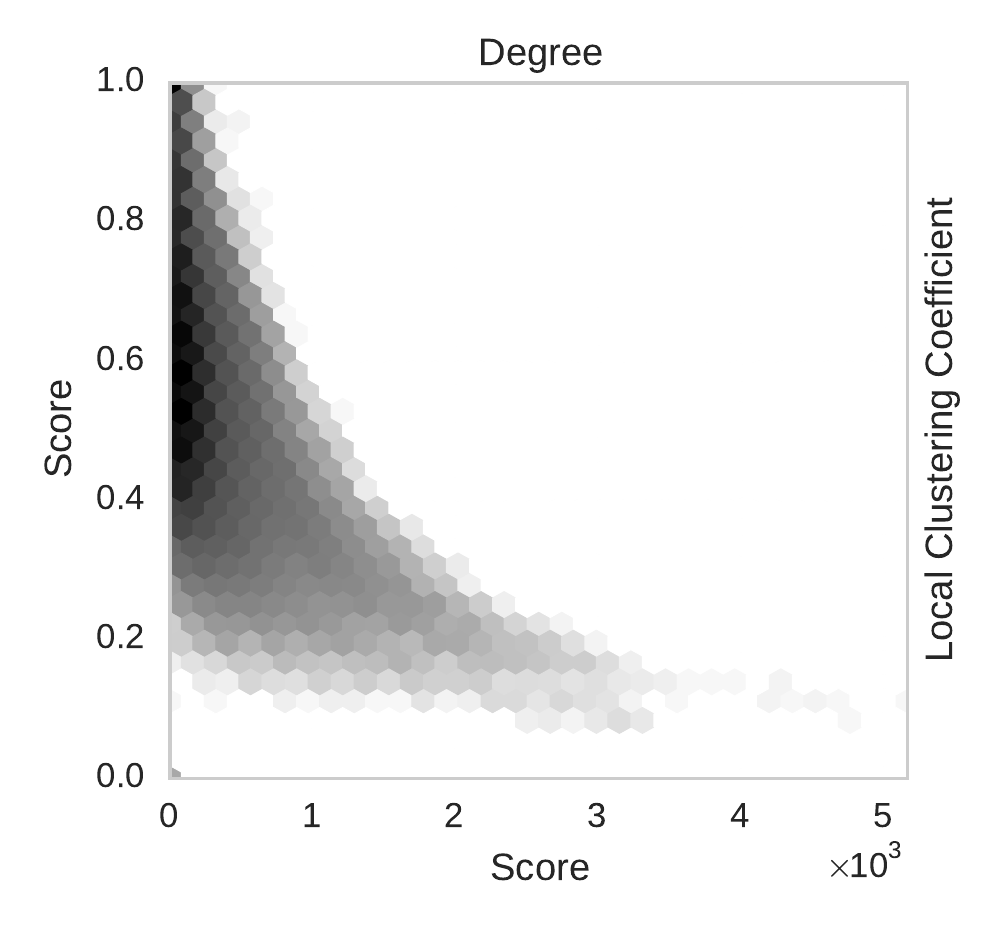}
	\caption{Scatter plot of degree and local clustering coefficient}
	\label{fig:connectome-scatter}
	\end{subfigure} 
	\qquad
	\begin{subfigure}[b]{0.5\textwidth}
	\includegraphics[width=\columnwidth]{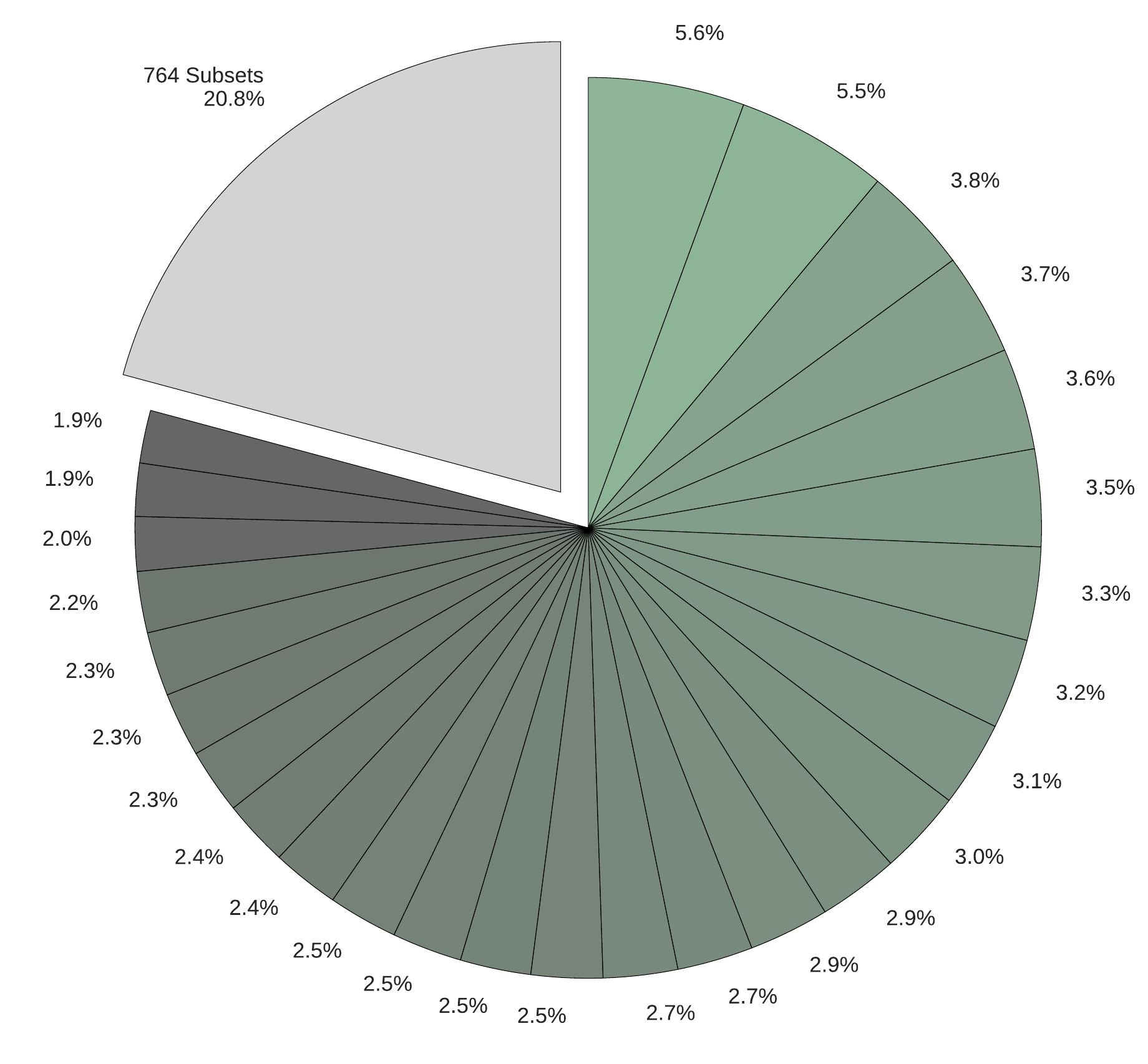}
	\caption{Size distribution of modularity-based communities}
	\label{fig:connectome-communities}
	\end{subfigure}
\end{center}
\caption{Statistical graphics from the profile of the connectome graph \textsf{con-fiber\_big} }
\end{figure}

Making the most of \nwk as a library requires writing some amount of custom code and some expertise in selecting algorithms and their parameters. This is one reason why we also provide an interface that makes exploratory analysis of large networks easy and fast even for non-expert users, and provides an extensive overview.
The underlying module assembles many algorithms into one program, automates analysis tasks and produces a graphical report to be displayed in the Jupyter Notebook or exported to an HTML or \LaTeX~report document.
Such a \textit{network profile} gives a statistical overview over the properties of the network.
It consists of the following parts: First global properties such as size and density are reported.
The report then focuses on a variety of node centrality measures, showing an overview of their distributions in the network (see Fig.~\ref{fig:measure-overview}).
Detailed views for centrality measures (see Fig.~\ref{fig:centrality-detail}) follow: Their distributions are plotted in histograms and characterized with standard statistics, and network-specific measures such as centralization and assortativity are shown.
We propose that correlations between centralities are per se interesting empirical features of a network. 
For instance, betweenness may or may not be positively correlated with increasing node degree. The prevalence of low-degree, high-betweenness nodes may influence the resilience of a transport network, as only few links then need to be severed in order to significantly disrupt transport processes following shortest paths. 
For the purpose of studying such aspects, the report displays a matrix of Spearman's correlation coefficients, showing how node ranks derived from the centrality measures correlate with each other (see Fig.~\ref{fig:connectome-correlations}). Furthermore, scatter plots for each combination of centrality measure are shown, suggesting the type of correlation (see Fig.~\ref{fig:connectome-scatter}).
The report continues with different ways of partitioning the network, showing histograms and pie charts for the
size distributions of connected components, modularity-based communities (see Fig.~\ref{fig:connectome-communities}) and $k$-shells, respectively.
Absent on purpose is a node-edge diagram of the graph, since graph drawing (apart from being computationally expensive) is usually not the preferred method to explore large complex networks. Rather, we consider networks first of all to be statistical data sets whose properties should be determined via graph algorithms and the results summarized via statistical graphics. 
The default configuration of the module is such that even networks with hundreds of millions of edges can be characterized in minutes on a parallel workstation.
Furthermore, it can be configured by the user depending on the desired choice of analytics and level of detail, so that custom reports can be generated. 

To pick an example from a scientific domain, the human connectome network \textsf{con-fiber\_big} maps brain regions and their anatomical connections at a relatively high resolution, yielding a graph with ca. 46 million edges. 
As the resolution of brain imaging technology improves, connectome analysis is likely to yield ever more massive network data sets, 
considering that the brain at the neuronal 
scale is a complex network on the order of $10^{10}$ nodes and $10^{14}$ edges.
On a first look, the network has properties similar to a social network, with a skewed degree distribution and high clustering.
The pattern of correlations (Fig.~\ref{fig:connectome-correlations}) differs from that of a typical friendship network (Fig.~\ref{fig:facebook-correlations}), with weaker positive correlations across the spectrum of centrality measures. As one observation to focus on, we may pick the strong negative correlation between the local clustering coefficient on the one hand and the PageRank and betweenness centrality on the other. 
High betweenness nodes are located on many shortest paths, and high PageRank results from being connected to neighbors which are themselves highly connected.
Thus, the correlations point to the presence of structural hub nodes that connect different brain regions which are not directly linked. 
Also, a look at a scatter plot generated (Fig.~\ref{fig:connectome-scatter}) reveals more details on the correlations: We see that the local clustering coefficient steadily falls with node degree, a majority of nodes having high clustering and low degree, a few nodes having low clustering and high degree.
Both observations are consistent with the finding of \textit{connector hub} regions situated along the midline of the brain, which are highly connected and link otherwise separated brain modules organized around smaller \textit{provincial hubs}~\cite{sporns2015modular}.

Another aspect we can focus on is community structure. There has been extensive research on the modular structure of brain networks, indicating that communities in the connectivity network can be interpreted as functional modules of the brain~\cite{sporns2015modular}. The communities found by the \textsf{PLM} modularity-maximizing heuristic in the \textsf{con-fiber\_big} graph can be interpreted accordingly. 
Their size distribution (Fig.~\ref{fig:connectome-communities}, in which a green pie slice represents the size of a community) shows that a large part of the network consists of about 30 communities of roughly equal size, in addition to a large number of very small communities (grey).
Of course, such interpretations of the network profile contain speculation, and a thorough analysis --  linking network structure to brain function -- would require the knowledge of a neuroscientist. 
Nonetheless, these examples illustrate how \nwk's capability to quickly generate an overview of structural properties can be used to generate hypotheses about the network data.

\section{Comparison to Related
Software}\label{sec:comparison}

Recent years have seen a proliferation of graph processing and network analysis software which vary widely in terms of target platform, user interface, scalability and feature set. 
We therefore locate \nwk relative to these efforts. 
Although the boundaries are not sharp, we would like to separate network analysis toolkits from general purpose graph frameworks  (\eg \href{http://www.boost.org/doc/libs/1_55_0/libs/graph/doc/table_of_contents.html}{Boost 
Graph Library} and \href{http://jung.sourceforge.net/}{\textsf{JUNG}}~\cite{o2003jung}), which are less focused on data analysis workflows.

As closest in terms of architecture, functionality and target use cases, we see
\href{http://igraph.sourceforge.net/}{\textsf{igraph}}~\cite{csardi2006igraph}
and \href{http://graph-tool.skewed.de/}{\textsf{graph-tool}}~\cite{graph-tool}.
They are packaged as Python modules, provide a broad feature set for network analysis workflows, and have active user communities.
\href{http://networkx.github.io/}{\textsf{NetworkX}}~\cite{hagberg2008exploring} is also a mature toolkit and the de-facto standard for the analysis of small to medium networks in
a Python environment, but not suitable for massive networks due to its pure Python implementations. 
(Due to the similar interface, users of \textsf{NetworkX} are likely to move easily to \nwk for larger networks.)
Like \nwk, \textsf{igraph} and \textsf{graph-tool} address the scalability issue by implementing core data structures and algorithms in C or C++.
\textsf{graph-tool} builds on the \textsf{Boost Graph Library} and parallelizes some kernels using \textsf{OpenMP}.
These similarities make those packages ideal candidates for an experimental comparison with \nwk (see Section~\ref{sub:comparative}).

Other projects are geared towards network science but differ in important aspects from \nwk.
\href{https://gephi.org/}{\textsf{Gephi}}~\cite{bastian2009gephi}, a GUI application for the Java platform, has a strong focus on visual network exploration.
\href{http://pajek.imfm.si/doku.php}{Pajek}~\cite{batagelj2004pajek}, a proprietary GUI application for the Windows operating system, also offers analysis capabilities similar to \nwk, as well as visualization features. The variant \textsf{PajekXXL} uses less memory and thus focuses on large datasets.

The \href{http://snap.stanford.edu/snappy/index.html}{\textsf{SNAP}}~\cite{snap} network analysis package has also recently adopted the hybrid approach of C++ core and Python interface. 
Related efforts from the algorithm engineering community are \href{http://kdt.sourceforge.net}{\textsf{KDT}}~\cite{kdt_sdm12} (built on an algebraic, distributed parallel backend), 
\href{http://trac.research.cc.gatech.edu/graphs/wiki/GraphCT}{\textsf{GraphCT}}~\cite{ediger2013graphct} (focused on massive multithreading architectures such as the \textsf{Cray XMT}),
\href{http://www.stingergraph.com/}{\textsf{STINGER}} (a dynamic graph data structure with some analysis capabilities)~\cite{6408680} and
\href{http://www.cs.cmu.edu/~jshun/ligra.html}{\textsf{Ligra}}~\cite{DBLP:conf/ppopp/ShunB13} (a recent shared-memory parallel library). 
They offer high performance through native, parallel implementations of certain kernels. 
However, to characterize a complex 
network in practice, we need a substantial set of analytics which those frameworks currently do not provide.

Among solutions for large-scale graph analytics, distributed computing frameworks (for instance \href{http://graphlab.org}{\textsf{GraphLab}}~\cite{low2012distributed}) are often prominently named.
However, graphs arising in many data analysis scenarios are not bigger than the billions of edges that fit into a conventional main memory and can therefore be processed far more efficiently in a shared-memory parallel model~\cite{DBLP:conf/ppopp/ShunB13}, which we confirm experimentally in a recent study~\cite{koch2105complex}. 
Distributed computing solutions become necessary 
for massive graph applications (as they appear, for example, in social media services), but we argue that shared-memory multicore machines go a long way for network science applications.

\section{Performance Evaluation}
\label{sec:performance}

\begin{table}[!h]
\centering
\begin{tabular}{|l|l|r|r|r|}

\rowcolor{gray!80}
\hline
\bf name                & \bf type & \bf $n$        & \bf $m$      & \bf source   \\
\hline
fb-Caltech36        &  social (friendship) & 769      & 16656   & \cite{traud2012social}   \\ 
PGPgiantcompo       & social (trust) & 10680    & 24316   & Bogu\~{n}a \etal 2014  \\ \rowcolor{bg}
coAuthorsDBLP       & coauthorship (science) & 299067   & 977676   & \cite{DBLP:reference/snam/BaderMS0KW14} \\
fb-Texas84          & social (friendship) & 36371    & 1590655 & \cite{traud2012social} \\ \rowcolor{bg}
Foursquare          & social (friendship) & 639014   & 3214986 & \cite{Zafarani+Liu:2009} \\
Lastfm              &  social (friendship) & 1193699  & 4519020 &  \cite{Zafarani+Liu:2009} \\ \rowcolor{bg}
wiki-Talk           & social  & 2394385  & 4659565 & \cite{snapnets}\\
Flickr              &  social (friendship) & 639014   & 55899882 &  \cite{Zafarani+Liu:2009} \\ \rowcolor{bg}
in-2004             &  web & 1382908  & 13591473  & \cite{BCSU3} \\
actor-collaboration & collaboration (film) & 382219   & 15038083 & \cite{kunegis2013konect}\\ \rowcolor{bg}
eu-2005             & web & 862664   & 16138468 &\cite{BCSU3} \\
flickr-growth-u     & social (friendship) & 2302925  & 33140017  & \cite{kunegis2013konect}\\ \rowcolor{bg}
con-fiber\_big      & brain (connectivity) & 591428   & 46374120 & http://openconnecto.me \\
Twitter             &  social (followership) & 15395404 & 85331845 &  \cite{Zafarani+Liu:2009}\\ \rowcolor{bg}
uk-2002             &  web & 18520486 & 261787258  &\cite{BCSU3}  \\ \hline
uk-2007-05 & web & 105896555 & 3301876564 & \cite{BCSU3}\\ \hline
\end{tabular}
\caption{Networks used in this paper}
\label{tab:networks}
\end{table}

This section presents an experimental evaluation of the performance of \nwk's algorithms.
Our platform is a shared-memory server with 256 GB RAM and 2x8 Intel(R) Xeon(R) E5-2680 cores (32 threads due to hyperthreading) at 2.7 GHz. 

\subsection{Benchmark}

\begin{figure}[!h]
\begin{center}
\includegraphics[width=.8\columnwidth]{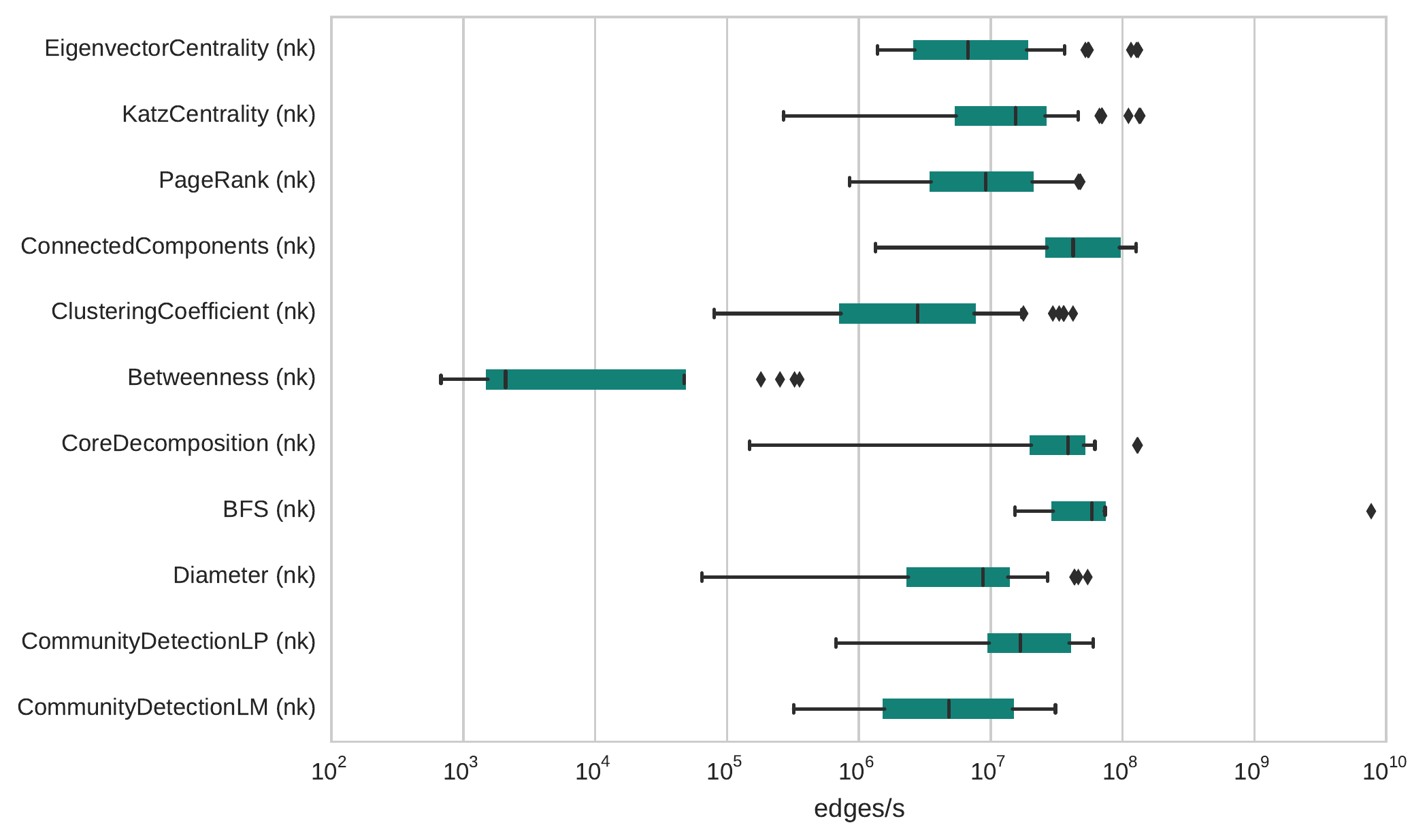}
\end{center}
\caption{Processing rates of \nwk analytics kernels} 
\label{fig:bench-main}
\end{figure}

\begin{figure}[!h]
\begin{center}
\includegraphics[width=0.8\columnwidth]{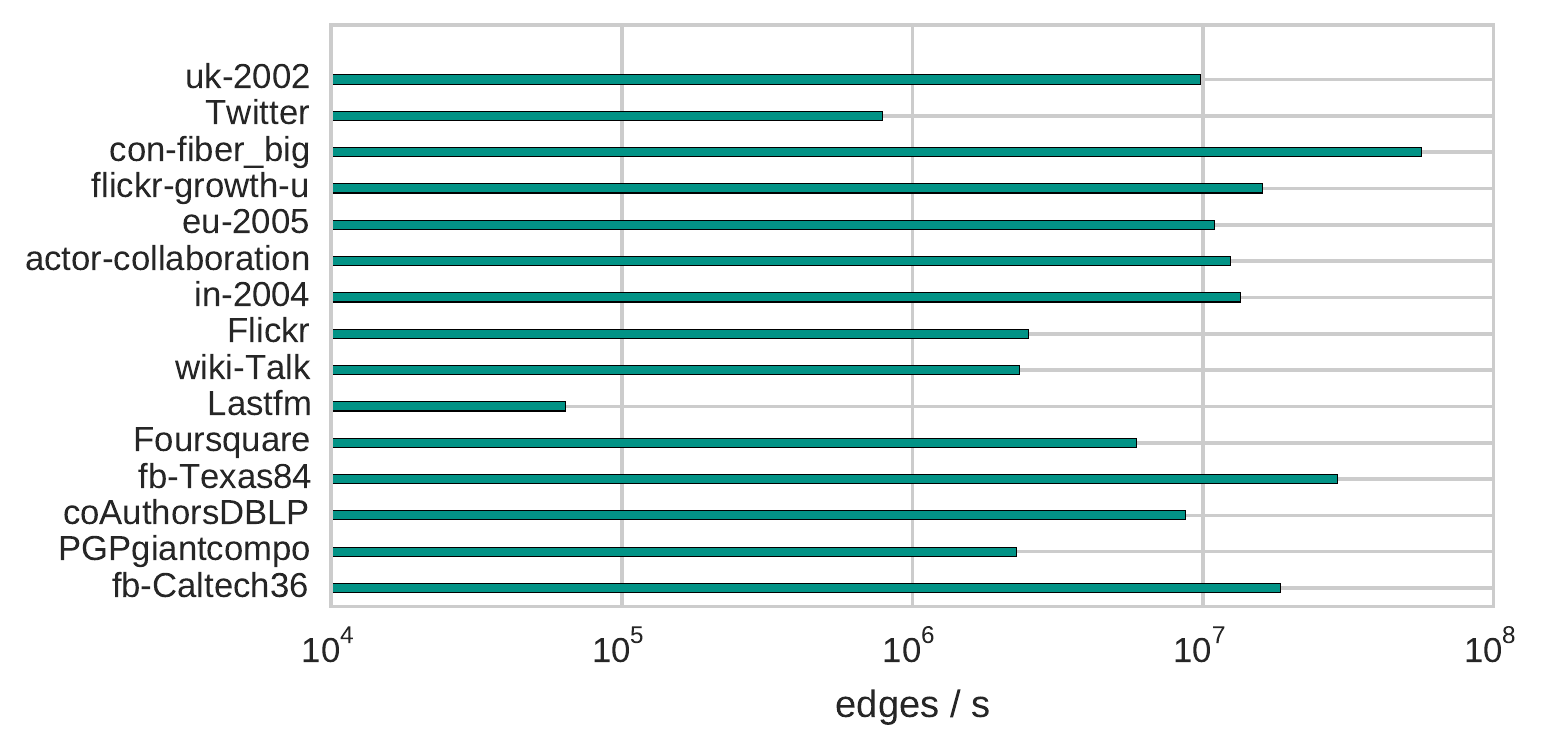}
\caption{Processing rate of diameter calculation on different networks}
\label{fig:bench-diameter}
\end{center}
\end{figure}

Fig.~\ref{fig:bench-main} shows results of a benchmark of the most important analytics kernels in \nwk.
The algorithms were applied to a diverse set of 15 real-world networks in the size range 
from 16k to 260M edges, including web graphs, social networks, connectome data and 
internet topology networks (see Table~\ref{tab:networks} for a description).
Kernels with quadratic running time (like \textsf{Betweenness}) were restricted to the subset of the 4 smallest networks.
The box plots illustrate the range of processing rates achieved (dots are outliers).
The benchmark illustrates that a set of efficient linear-time kernels, including \textsf{ConnectedComponents}, the community detectors, \textsf{PageRank}, \textsf{CoreDecomposition} and \textsf{ClusteringCoefficient},
scales well to networks in the order of $10^8$ edges.
The \textsf{iFub}~\cite{crescenzi2013computing} algorithm demonstrates its surprisingly good performance on complex networks, moving diameter calculation effectively into the class of linear-time kernels. Fig.~\ref{fig:bench-diameter} breaks its processing rate down to the particular instances, in decreasing order of size,
illustrating that performance is often strongly dependent on the specific structure of complex networks.
Algorithms like \textsf{BFS} and \textsf{ConnectedComponents} actually scan every edge at a rate of $10^7$ to $10^8$ edges per second.
Betweenness calculation remains very time-consuming in spite of parallelization, but approximate results can be obtained two order of magnitudes faster.

\subsection{Comparative Benchmark.}
\label{sub:comparative}

\begin{figure}[!h]
\begin{center}
\includegraphics[width=.8\columnwidth]{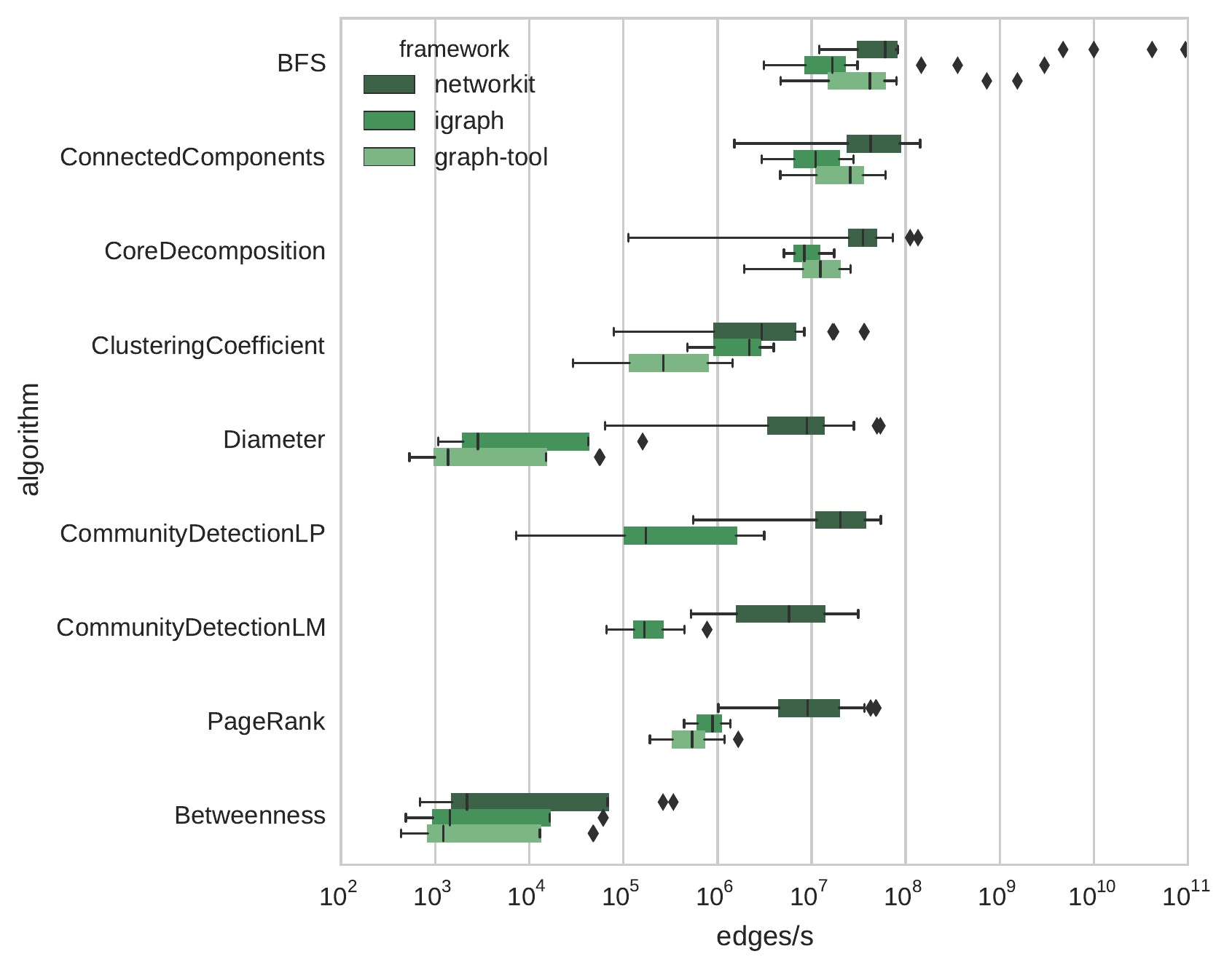}
\end{center}
\caption{Processing rates of typical analytics tasks: \nwk in comparison with \textsf{igraph} and \textsf{graph-tool} }
\label{fig:comparative-bench}
\end{figure}

\nwk, \textsf{igraph} and \textsf{graph-tool} rely on the same hybrid architecture of C/C++ implementations with a Python interface.
\textsf{igraph} uses non-parallel C code while \textsf{graph-tool} also features parallelism.
We benchmarked typical analysis kernels for the three packages in comparison on the aforementioned parallel platform and present the measured performance in Fig.~\ref{fig:comparative-bench}. 
Where applicable, algorithm parameters were selected to ensure a fair comparison. In this respect it should be mentioned that \textsf{graph-tool}'s implementation of Brandes' betweenness algorithm does more work as it also calculates edge betweenness scores during the run. (Anyway, performance differences in the implementation quickly become irrelevant for a non-linear-time algorithm as the input size grows.) \textsf{graph-tool} also takes a different approach to community detection, hence the comparison is between \textsf{igraph} and \nwk only.
We summarize the benchmark results as follows: In our benchmark, \nwk was the only framework that could consistently run the set of kernels (excluding the quadratic-time betweenness) on the full set of networks in the timeframe of an overnight run. For some of \textsf{igraph}'s and \textsf{graph-tool}'s implementations the test set had to be restricted to a subset of smaller networks to make it possible to run the complete benchmark over night. \nwk has the fastest average processing rate on all of these typical analytics kernels. Our implementations have a slight edge over the others for breadth-first search, connected components, clustering coefficients and betweenness. Considering that the algorithms are very similar, this is likely due to subtle differences and optimizations in the implementation. For PageRank, core decomposition and the two community detection algorithms, our parallel methods lead to a larger speed advantage. The massive difference for the diameter calculation is due to our choice of the \textsf{iFub} algorithm~\cite{crescenzi2013computing}, which has better running time in the typical case (\ie complex networks with hub structure) and enables the processing of inputs that are orders of magnitudes larger.

Another scalability factor is the memory footprint of the graph data structure. \nwk provides a lean implementation in which the 260M edges of the \texttt{uk-2002} web graph occupy only 9 GB, compared with \textsf{igraph} (93GB) and \textsf{graph-tool} (14GB).
After indexing the edges, \eg in order to compute edge centrality scores, \nwk requires 11 GB for the graph.

A third factor that should not be ignored for real workflows is I/O. Getting a large graph from hard disk to memory often takes far longer than the actual analysis. For our benchmark, we chose the GML graph file format for the input files, because it is supported by all three frameworks. We observed that the \nwk parser is significantly faster for these non-attributed graphs. 

\begin{figure}[!h]
\begin{center}
\includegraphics[width=.75\columnwidth]{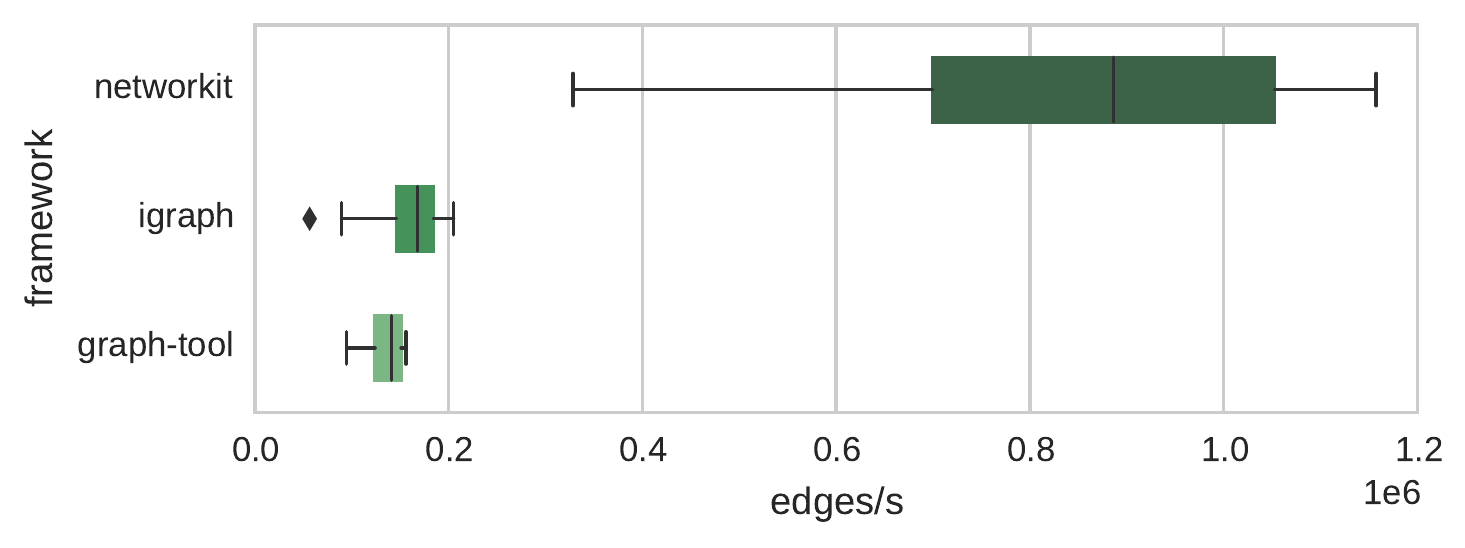}
\end{center}
\caption{I/O rates of reading a graph from a GML file: \nwk in comparison with \textsf{igraph} and \textsf{graph-tool} }
\label{fig:comparative-read}
\end{figure}


%
\section{Open-Source Development and Distribution}

Through open-source development we would like to encourage usage and contributions by a diverse community, including data mining users and algorithm engineers. 
While the core developer team is located at KIT, \nwk is becoming a community project with a growing number of external users and contributors.
The code is free software licensed under the permissive MIT License.
The package source, documentation, and additional resources can be obtained from \url{http://networkit.iti.kit.edu}.
The package \texttt{networkit} is also installable via the Python package manager \texttt{pip}.
For custom-built applications, the Python layer may be omitted by building a subset of functionality as a native library.

\section{Conclusion}

The \nwk project exists at the intersection of graph algorithm research and network science.
Its contributors develop and collect state-of-the-art algorithms for network analysis tasks and incorporate them into ready-to-use software. The open-source package is under continuous development.
The result is a tool suite of network analytics kernels, network generators and utility software to
explore and characterize large network data sets on typical multicore computers.
We detailed techniques that allow \nwk to scale to large networks, including appropriate algorithm patterns (parallelism, heuristics, data structures) and implementation patterns (\eg modular design).
The interface provided by our Python module allows domain experts to focus on data analysis workflows instead of the intricacies of programming. 
This is especially enabled by a new frontend that generates comprehensive statistical reports on structural features of the network.
 Specialized programming skills are not required, though users familiar with the Python ecosystem of data analysis tools will appreciate the possibility to seamlessly integrate our toolkit.

Among similar software packages, \nwk yields the best performance for common analysis workflows.
Our experimental study showed that \nwk is capable of quickly processing large-scale networks for a variety of analytics kernels in a reliable manner. This translates into faster workflow and extended analysis capabilities in practice.
We recommend \nwk for the comprehensive structural analysis of large complex networks, as well as processing large batches of smaller networks. With fast parallel algorithms, scalability is in practice primarily limited by the size of the shared memory: A standard multicore workstation with 256 GB RAM can therefore process up to $10^{10}$ edge graphs.

\subsubsection*{{\scriptsize{Acknowledgements.}}}

\begin{scriptsize}
This work was partially supported by the project \emph{Parallel
Analysis of Dynamic Networks -- Algorithm Engineering of
Efficient Combinatorial and Numerical Methods}, which
is funded by the Ministry of Science, Research and the Arts Baden-W\"urttemberg and by
DFG grant ME-3619/3-1 (FINCA) within the SPP 1736 \emph{Algorithms for Big Data}.
Aleksejs Sazonovs acknowledges support by the RISE program of the German Academic Exchange Service (DAAD).
We thank Maximilian Vogel and Michael Hamann for continuous algorithm and software engineering work on the package. We also thank 
Lukas Barth, Miriam Beddig, Elisabetta Bergamini, Stefan Bertsch, Pratistha Bhattarai, Andreas Bilke, Simon Bischof, Guido Br\"{u}ckner, Mark Erb, Kolja Esders, Patrick Flick, Lukas Hartmann, Daniel Hoske, Gerd Lindner, Moritz v. Looz, Yassine Marrakchi, Mustafa \"{O}zdayi, Marcel Radermacher, Klara Reichard, Matteo Riondato, Marvin Ritter, Arie Slobbe, Florian Weber, Michael Wegner and J\"{o}rg Weisbarth for contributing to the project.

\end{scriptsize}


\bibliographystyle{apalike}
\bibliography{Bibliography}

\begin{thebibliography}{}

\bibitem[Aiello et~al., 2000]{aiello2000random}
Aiello, W., Chung, F., and Lu, L. (2000).
\newblock A random graph model for massive graphs.
\newblock In {\em Proceedings of the thirty-second annual ACM symposium on
  Theory of computing}, pages 171--180. Acm.

\bibitem[Alahakoon et~al., 2011]{alahakoon2011k}
Alahakoon, T., Tripathi, R., Kourtellis, N., Simha, R., and Iamnitchi, A.
  (2011).
\newblock K-path centrality: A new centrality measure in social networks.
\newblock In {\em Proceedings of the 4th Workshop on Social Network Systems},
  page~1. ACM.

\bibitem[Albert and Barab{\'a}si, 2002]{albert2002statistical}
Albert, R. and Barab{\'a}si, A. (2002).
\newblock Statistical mechanics of complex networks.
\newblock {\em Reviews of modern physics}, 74(1):47.

\bibitem[Bader et~al., 2014]{DBLP:reference/snam/BaderMS0KW14}
Bader, D.~A., Meyerhenke, H., Sanders, P., Schulz, C., Kappes, A., and Wagner,
  D. (2014).
\newblock Benchmarking for graph clustering and partitioning.
\newblock In {\em Encyclopedia of Social Network Analysis and Mining}, pages
  73--82.

\bibitem[Bastian et~al., 2009]{bastian2009gephi}
Bastian, M., Heymann, S., and Jacomy, M. (2009).
\newblock Gephi: an open source software for exploring and manipulating
  networks.
\newblock In {\em International Conference on Weblogs and Social Media}, pages
  361--362.

\bibitem[Batagelj and Brandes, 2005]{batagelj2005efficient}
Batagelj, V. and Brandes, U. (2005).
\newblock Efficient generation of large random networks.
\newblock {\em Physical Review E}, 71(3):036113.

\bibitem[Batagelj and Mrvar, 2004]{batagelj2004pajek}
Batagelj, V. and Mrvar, A. (2004).
\newblock {\em Pajek---analysis and visualization of large networks}, volume
  2265 of the series Lecture Notes in Computer Science pp 477-478.
\newblock Springer.

\bibitem[Batagelj and Zaver{\v s}nik, 2011]{Batagelj:2011fk}
Batagelj, V. and Zaver{\v s}nik, M. (2011).
\newblock Fast algorithms for determining (generalized) core groups in social
  networks.
\newblock {\em Advances in Data Analysis and Classification}, 5(2):129--145.

\bibitem[Behnel et~al., 2011]{behnel2011cython}
Behnel, S., Bradshaw, R., Citro, C., Dalcin, L., Seljebotn, D.~S., and Smith,
  K. (2011).
\newblock Cython: The best of both worlds.
\newblock {\em Computing in Science \& Engineering}, 13(2):31--39.

\bibitem[Bergamini and Meyerhenke, 2015]{DBLP:conf/esa/BergaminiM15}
Bergamini, E. and Meyerhenke, H. (2015).
\newblock Fully-dynamic approximation of betweenness centrality.
\newblock In {\em Algorithms - {ESA} 2015 - 23rd Annual European Symposium,
  Patras, Greece, September 14-16, 2015, Proceedings}, pages 155--166.

\bibitem[Bergamini et~al., 2015]{DBLP:conf/alenex/BergaminiMS15}
Bergamini, E., Meyerhenke, H., and Staudt, C. (2015).
\newblock Approximating betweenness centrality in large evolving networks.
\newblock In {\em Proceedings of the Seventeenth Workshop on Algorithm
  Engineering and Experiments, {ALENEX} 2015, San Diego, CA, USA, January 5,
  2015}, pages 133--146.

\bibitem[Blondel et~al., 2008]{Blondel:2008uq}
Blondel, V.~D., Guillaume, J.-L., Lambiotte, R., and Lefebvre, E. (2008).
\newblock Fast unfolding of communities in large networks.
\newblock {\em Journal of Statistical Mechanics: Theory and Experiment},
  2008(10):P10008.

\bibitem[Boccaletti et~al., 2006]{boccaletti2006complex}
Boccaletti, S., Latora, V., Moreno, Y., Chavez, M., and Hwang, D.-U. (2006).
\newblock Complex networks: Structure and dynamics.
\newblock {\em Physics reports}, 424(4):175--308.

\bibitem[Bode et~al., 2014]{bode2014probability}
Bode, M., Fountoulakis, N., and M{\"u}ller, T. (2014).
\newblock The probability that the hyperbolic random graph is connected.
\newblock Preprint available at
  \url{http://www.staff.science.uu.nl/~muell001/Papers/BFM.pdf}.

\bibitem[Boldi et~al., 2004]{BCSU3}
Boldi, P., Codenotti, B., Santini, M., and Vigna, S. (2004).
\newblock Ubicrawler: A scalable fully distributed web crawler.
\newblock {\em Software: Practice \& Experience}, 34(8):711--726.

\bibitem[Brandes, 2001]{Bra01}
Brandes, U. (2001).
\newblock A faster algorithm for betweenness centrality.
\newblock {\em J.\ Mathematical Sociology}, 25(2):163--177.

\bibitem[CAIDA, 2003]{caida}
CAIDA (2003).
\newblock Caida skitter router-level topology measurements.
\newblock \url{http://www.caida.org/data/router-adjacencies/}.

\bibitem[Chakrabarti et~al., 2004]{chakrabarti2004r}
Chakrabarti, D., Zhan, Y., and Faloutsos, C. (2004).
\newblock {R-MAT: A recursive model for graph mining}.
\newblock {\em Computer Science Department}, page 541.

\bibitem[Costa et~al., 2011]{costa2011analyzing}
Costa, L. d.~F., Oliveira~Jr, O.~N., Travieso, G., Rodrigues, F.~A.,
  Villas~Boas, P.~R., Antiqueira, L., Viana, M.~P., and Correa~Rocha, L.~E.
  (2011).
\newblock Analyzing and modeling real-world phenomena with complex networks: a
  survey of applications.
\newblock {\em Advances in Physics}, 60(3):329--412.

\bibitem[Crescenzi et~al., 2013]{crescenzi2013computing}
Crescenzi, P., Grossi, R., Habib, M., Lanzi, L., and Marino, A. (2013).
\newblock On computing the diameter of real-world undirected graphs.
\newblock {\em Theoretical Computer Science}, 514:84--95.

\bibitem[Csardi and Nepusz, 2006]{csardi2006igraph}
Csardi, G. and Nepusz, T. (2006).
\newblock The igraph software package for complex network research.
\newblock {\em InterJournal, Complex Systems}, 1695(5).

\bibitem[Dasari et~al., 2014]{DBLP:conf/bigdataconf/DasariRZ14}
Dasari, N.~S., Ranjan, D., and Zubair, M. (2014).
\newblock {ParK: An efficient algorithm for k-core decomposition on multicore
  processors}.
\newblock In Lin, J., Pei, J., Hu, X., Chang, W., Nambiar, R., Aggarwal, C.,
  Cercone, N., Honavar, V., Huan, J., Mobasher, B., and Pyne, S., editors, {\em
  2014 {IEEE} International Conference on Big Data, Big Data 2014, Washington,
  DC, USA, October 27-30, 2014}, pages 9--16. {IEEE}.

\bibitem[Ediger et~al., 2013]{ediger2013graphct}
Ediger, D., Jiang, K., Riedy, E.~J., and Bader, D.~A. (2013).
\newblock {GraphCT: Multithreaded algorithms for massive graph analysis}.
\newblock {\em Parallel and Distributed Systems, IEEE Transactions on},
  24(11):2220--2229.

\bibitem[Ediger et~al., 2012]{6408680}
Ediger, D., McColl, R., Riedy, J., and Bader, D. (2012).
\newblock {STINGER: High performance data structure for streaming graphs}.
\newblock In {\em High Performance Extreme Computing (HPEC), 2012 IEEE
  Conference on}, pages 1--5.

\bibitem[Eppstein and Wang, 2004]{eppstein2004fast}
Eppstein, D. and Wang, J. (2004).
\newblock Fast approximation of centrality.
\newblock {\em J. Graph Algorithms Appl.}, 8:39--45.

\bibitem[Esders, 2015]{esders2015linkprediction}
Esders, K. (2015).
\newblock Link prediction in large-scale complex networks.
\newblock Master's thesis, Karlsruhe Institute of Technology,
  \url{http://parco.iti.kit.edu/attachments/Kolja%20Esders%20-%20Thesis.pdf}.

\bibitem[Flick, 2014]{flickanalysis}
Flick, P. (2014).
\newblock Analysis of human tissue-specific protein-protein interaction
  networks.
\newblock Master's thesis, Karlsruhe Institute of Technology.

\bibitem[Freeman, 1979]{freeman1979centrality}
Freeman, L.~C. (1979).
\newblock Centrality in social networks conceptual clarification.
\newblock {\em Social networks}, 1(3):215--239.

\bibitem[Geisberger et~al., 2008]{geisberger2008better}
Geisberger, R., Sanders, P., and Schultes, D. (2008).
\newblock Better approximation of betweenness centrality.
\newblock In {\em ALENEX}, pages 90--100. SIAM.

\bibitem[Girvan and Newman, 2002]{girvan2002community}
Girvan, M. and Newman, M. (2002).
\newblock Community structure in social and biological networks.
\newblock {\em Proc. of the National Academy of Sciences}, 99(12):7821.

\bibitem[Gugelmann et~al., 2012]{DBLP:conf/icalp/GugelmannPP12}
Gugelmann, L., Panagiotou, K., and Peter, U. (2012).
\newblock Random hyperbolic graphs: Degree sequence and clustering - (extended
  abstract).
\newblock In {\em Automata, Languages, and Programming - 39th International
  Colloquium, {ICALP} 2012, Proceedings, Part {II}}, pages 573--585.

\bibitem[Hagberg et~al., 2008]{hagberg2008exploring}
Hagberg, A., Swart, P., and S~Chult, D. (2008).
\newblock {Exploring network structure, dynamics, and function using NetworkX}.
\newblock Technical report, Los Alamos National Laboratory (LANL).

\bibitem[Hakimi, 1962]{hakimi1962realizability}
Hakimi, S.~L. (1962).
\newblock On realizability of a set of integers as degrees of the vertices of a
  linear graph. i.
\newblock {\em Journal of the Society for Industrial \& Applied Mathematics},
  10(3):496--506.

\bibitem[Katz, 1953]{katz1953new}
Katz, L. (1953).
\newblock A new status index derived from sociometric analysis.
\newblock {\em Psychometrika}, 18(1):39--43.

\bibitem[Kiwi and Mitsche, 2015]{kiwi2015bound}
Kiwi, M. and Mitsche, D. (2015).
\newblock A bound for the diameter of random hyperbolic graphs.
\newblock {\em preprint available at http://arxiv. org/abs/1408.2947}.

\bibitem[Koch et~al., 2015]{koch2105complex}
Koch, J., Staudt, C.~L., Vogel, M., and Meyerhenke, H. (2015).
\newblock Complex network analysis on distributed systems: An empirical
  comparison.
\newblock In {\em International Symposium on Foundations and Applications of
  Big Data Analytics}.

\bibitem[Krioukov et~al., 2010]{Krioukov2010}
Krioukov, D., Papadopoulos, F., Kitsak, M., Vahdat, A., and Bogu\~n\'a, M.
  (2010).
\newblock Hyperbolic geometry of complex networks.
\newblock {\em Physical Review E}, 82:036106.

\bibitem[Kunegis, 2013]{kunegis2013konect}
Kunegis, J. (2013).
\newblock Konect: the koblenz network collection.
\newblock In {\em Proceedings of the 22nd international conference on World
  Wide Web companion}, pages 1343--1350. International World Wide Web
  Conferences Steering Committee.

\bibitem[Lancichinetti and Fortunato, 2009]{lancichinetti2009benchmarks}
Lancichinetti, A. and Fortunato, S. (2009).
\newblock Benchmarks for testing community detection algorithms on directed and
  weighted graphs with overlapping communities.
\newblock {\em Physical Review E}, 80(1):016118.

\bibitem[Leskovec and Krevl, 2014]{snapnets}
Leskovec, J. and Krevl, A. (2014).
\newblock {SNAP Datasets}: {Stanford} large network dataset collection.
\newblock \url{http://snap.stanford.edu/data}.

\bibitem[Leskovec and Sosi\v{c}, 2014]{snap}
Leskovec, J. and Sosi\v{c}, R. (2014).
\newblock {SNAP}: A general purpose network analysis and graph mining library
  in {C++}.
\newblock \url{http://snap.stanford.edu/snap}.

\bibitem[Lindner et~al., 2015]{lindner2015structure}
Lindner, G., Staudt, C.~L., Hamann, M., Meyerhenke, H., and Wagner, D. (2015).
\newblock Structure-preserving sparsification of social networks.
\newblock {\em ASONAM}.

\bibitem[Low et~al., 2012]{low2012distributed}
Low, Y., Bickson, D., Gonzalez, J., Guestrin, C., Kyrola, A., and Hellerstein,
  J.~M. (2012).
\newblock Distributed graphlab: a framework for machine learning and data
  mining in the cloud.
\newblock {\em Proceedings of the VLDB Endowment}, 5(8):716--727.

\bibitem[Lugowski et~al., 2012]{kdt_sdm12}
Lugowski, A., Alber, D., Bulu\c{c}, A., Gilbert, J.~R., Reinhardt, S., Teng,
  Y., and Waranis, A. (2012).
\newblock A flexible open-source toolbox for scalable complex graph analysis.
\newblock In {\em Proceedings of the Twelfth SIAM International Conference on
  Data Mining (SDM12)}, pages 930--941.

\bibitem[Newman, 2010]{newman2010networks}
Newman, M. (2010).
\newblock {\em Networks: an introduction}.
\newblock Oxford University Press.

\bibitem[O'Madadhain et~al., 2003]{o2003jung}
O'Madadhain, J., Fisher, D., White, S., and Boey, Y. (2003).
\newblock The {JUNG} (java universal network/graph) framework.
\newblock {\em University of California, Irvine, California}.

\bibitem[{P. Erd{\H o}s}, 1960]{P.Erdos}
{P. Erd{\H o}s}, A.~R. (1960).
\newblock {On the Evolution of Random Graphs}.
\newblock {\em Publication of the Mathematical Institute of the Hungarian
  Academy of Sciences}.

\bibitem[Page et~al., 1999]{page1999pagerank}
Page, L., Brin, S., Motwani, R., and Winograd, T. (1999).
\newblock The pagerank citation ranking: Bringing order to the web.

\bibitem[Peixoto, 2015]{graph-tool}
Peixoto, T.~P. (2015).
\newblock graph-tool.
\newblock \url{http://graph-tool.skewed.de}.

\bibitem[Perez et~al., 2013]{perez2013open}
Perez, F., Granger, B.~E., and Obispo, C. (2013).
\newblock An open source framework for interactive, collaborative and
  reproducible scientific computing and education.

\bibitem[Raghavan et~al., 2007]{Raghavan:2007fk}
Raghavan, U.~N., Albert, R., and Kumara, S. (2007).
\newblock Near linear time algorithm to detect community structures in
  large-scale networks.
\newblock {\em Physical Review E}, 76(3):036106.

\bibitem[Riondato and Kornaropoulos, 2015]{Riondato:2015}
Riondato, M. and Kornaropoulos, E. (2015).
\newblock Fast approximation of betweenness centrality through sampling.
\newblock {\em Data Mining and Knowledge Discovery}, pages 1--38.

\bibitem[Schank and Wagner, 2005]{schank2005approximating}
Schank, T. and Wagner, D. (2005).
\newblock Approximating clustering coefficient and transitivity.
\newblock {\em Journal of Graph Algorithms and Applications}, 9(2):265--275.

\bibitem[Shun and Blelloch, 2013]{DBLP:conf/ppopp/ShunB13}
Shun, J. and Blelloch, G.~E. (2013).
\newblock Ligra: a lightweight graph processing framework for shared memory.
\newblock In {\em {ACM} {SIGPLAN} Symposium on Principles and Practice of
  Parallel Programming, PPoPP '13, Shenzhen, China, February 23-27, 2013},
  pages 135--146.

\bibitem[Sporns and Betzel, 2015]{sporns2015modular}
Sporns, O. and Betzel, R.~F. (2015).
\newblock Modular brain networks.
\newblock {\em Annual review of psychology}, 67(1).

\bibitem[Staudt and Meyerhenke, 2015]{staudt2015engineering}
Staudt, C. and Meyerhenke, H. (2015).
\newblock Engineering parallel algorithms for community detection in massive
  networks.
\newblock {\em Parallel and Distributed Systems, IEEE Transactions on},
  PP(99):1--1.

\bibitem[Traud et~al., 2012]{traud2012social}
Traud, A.~L., Mucha, P.~J., and Porter, M.~A. (2012).
\newblock Social structure of facebook networks.
\newblock {\em Physica A: Statistical Mechanics and its Applications},
  391(16):4165--4180.

\bibitem[von Looz and Meyerhenke, 2015]{von2015querying}
von Looz, M. and Meyerhenke, H. (2015).
\newblock Querying probabilistic neighborhoods in spatial data sets
  efficiently.
\newblock {\em arXiv preprint arXiv:1509.01990}.

\bibitem[von Looz et~al., 2015]{LoozMP15generating}
von Looz, M., Meyerhenke, H., and Prutkin, R. (2015).
\newblock Generating random hyperbolic graphs in subquadratic time.
\newblock In {\em Proc.\ 26th Int'l Symp.\ on Algorithms and Computation (ISAAC
  2015)}, LNCS. Springer.
\newblock To appear.

\bibitem[Zafarani and Liu, 2009]{Zafarani+Liu:2009}
Zafarani, R. and Liu, H. (2009).
\newblock Social computing data repository at {ASU}.
\newblock \url{http://socialcomputing.asu.edu}.

\end{thebibliography}

\end{document}